\documentclass[a4paper]{article}

\usepackage{textcomp}
\usepackage[mathlines]{lineno} % for line number
\usepackage{booktabs}
\usepackage{float}
\restylefloat{table}
\usepackage{rotating}
\usepackage{graphicx, subfigure}
\usepackage{setspace}
\usepackage[font={small,it}]{caption}
\usepackage{epsf}
\usepackage{bm}
\usepackage{latexsym}
\usepackage{enumerate}
\usepackage{framed}
\usepackage{xcolor}
\usepackage[round,comma,authoryear]{natbib}
\usepackage{amsmath, amssymb, amsfonts}
\usepackage{epsfig}
\usepackage{graphicx}
\RequirePackage{amsopn}
\RequirePackage{amsfonts}
\RequirePackage{amsthm}
\usepackage{amsmath, amssymb, amsfonts}
\usepackage[english]{babel}
\usepackage{graphicx}
\usepackage{algorithm}
\usepackage{algorithmic}
\usepackage{multirow}
\usepackage{todonotes}
\usepackage{color}
\usepackage{hyperref}
\graphicspath{{./figs/}}

\renewcommand\appendix{\par
  \setcounter{section}{0}
  \setcounter{subsection}{0}
  \setcounter{figure}{0}
  \setcounter{table}{0}
  \renewcommand\thesection{Appendix \Alph{section}}
  \renewcommand\thefigure{\Alph{section}\arabic{figure}}
  \renewcommand\thetable{\Alph{section}\arabic{table}}
}

\newtheorem{theorem}{Theorem}[section]

\newcommand{\Ehat}{\overline{E}}

\title{Bayesian Gaussian Copula Graphical Modeling for Dupuytren Disease}
\author{
  Abdolreza Mohammadi       \\ University of Groningen            \\ \texttt{a.mohammadi@rug.nl}
  \and Fentaw Abegaz        \\ University of Groningen            \\ \texttt{f.abegaz.yazew@rug.nl}
  \and Edwin van den Heuvel \\ Eindhoven University of Technology \\ \texttt{e.r.v.d.heuvel@tue.nl}
  \and Ernst C. Wit         \\ University of Groningen            \\ \texttt{e.c.wit@rug.nl}
  }
\begin{document}

\maketitle

\begin{abstract}
Dupuytren disease is a fibroproliferative disorder with unknown etiology that often progresses and eventually can cause permanent contractures of the affected fingers. 
% Most of the research on severity of the disease and the phenotype of this disease are observational studies without concrete statistical analyses.
% There is a lack of multivariate analysis for the disease taking into account potential risk factors.
In this paper, we provide a computationally efficient Bayesian framework to discover potential risk factors and investigate which fingers are jointly affected.
Our Bayesian approach is based on Gaussian copula graphical models, 
which are one potential way to discover the underlying conditional independence structure of variables in multivariate mixed data.
In particular, we combine the semiparametric Gaussian copula with extended rank likelihood which is appropriate to analyse multivariate mixed data with arbitrary marginal distributions.
For the graph structure learning, we construct a computationally efficient search algorithm which is a trans-dimensional MCMC algorithm based on a birth-death process. % with an appropriate stationary distribution.
In addition, to make our statistical method easily accessible to other researchers, we have implemented our method in {\tt C++} and interfaced with {\tt R} software as an {\tt R} package 
{\tt BDgraph} which is available at \url{http://CRAN.R-project.org/package=BDgraph}.

\textbf{Keywords}: Dupuytren disease; Risk factors; Bayesian inference; Gaussian copula graphical models; Bayesian model selection; Latent variable models; 
                   Birth-death process; Markov chain Monte Carlo.
\end{abstract}

%%%%%%%%%%%%%%%%%%%%%%%%%%%%%%%%%%%%%%%%%%%%%%%%%%%%%%%%%%%%%%%%%%%%%%%%%%%%%%%%%%%%%%%%%%%%%%%%%%%%%%%%%%%%%%%%%%%%%%%%%%%%%%%%%%%%%%%%%%%%%%%%%%%%%%%%%%%%%%%%%%%%%
\section{Introduction}
\label{sec:intro}

\begin{figure}[ht]
\centering
\subfigure[]{
   \centering
   \includegraphics[width=0.29\textwidth]{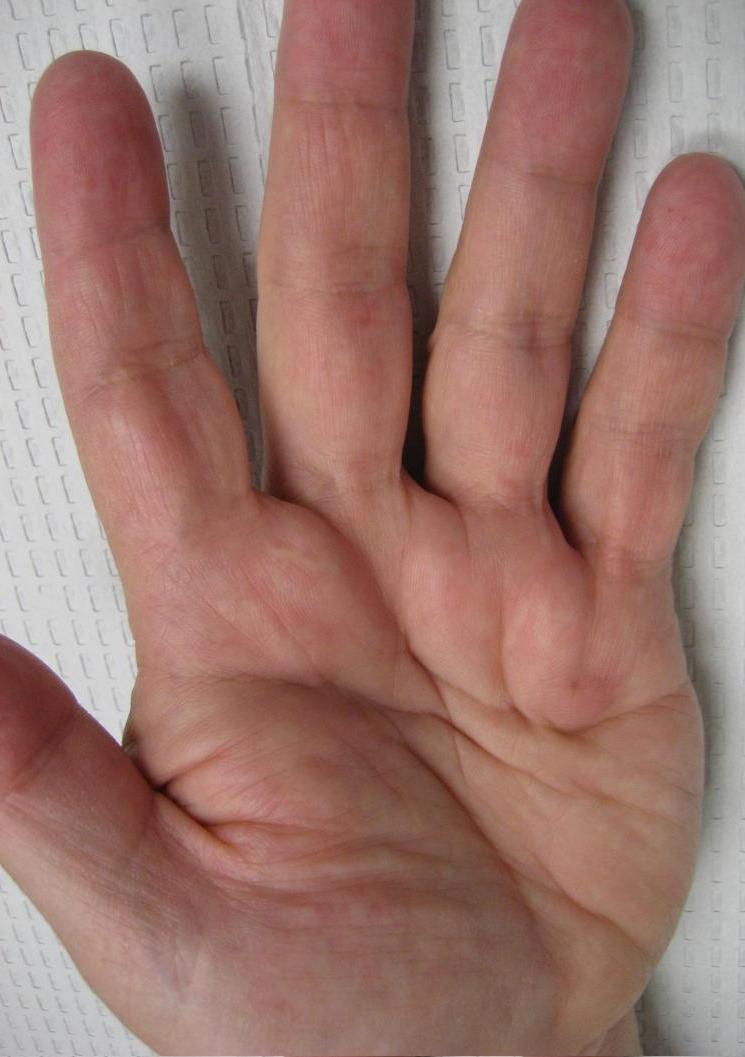}
   \label{fig:hand image-1}
}
\subfigure[]{
   \centering
    \includegraphics[width=0.29\textwidth]{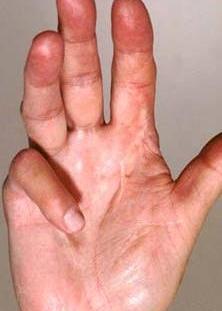}
    \label{fig:hand image-2}
}
\caption[]{(a) hand image of a patient with Dupuytren disease whose fingers have not been affected by the disease. 
	   Palmar nodules and small cords have no signs of contracture.
           (b) hand image of a patient with Dupuytren disease whose finger have been affected by the disease.}
\label{fig:hand image}
\end{figure}

Dupuytren disease is a hereditary disorder that is present worldwide.
It is however more prevalent in people with northern European ancestry \citep{bayat2006management}.
The disease is an incurable fibroproliferative disorder that alters the palmar fascia of the hand and may causes progressive and permanent flexion contracture of the fingers.
Initially, skin pittings and subcutaneous nodules appear in the palm; see Figure \ref{fig:hand image-1}.
At a later stage, cords appear that connect the nodules and may contract the fingers into a flexed position; see Figure \ref{fig:hand image-2}. 
Contracture can arise in a single ray or in multiple rays. 
The disease mostly appears on the ulnar side of the hand, i.e., it affects the pink and ring fingers most-frequently (see Figure \ref{fig:plot data}). 
The only available treatment is surgical intervention.
Although much is known about the disease, the questions arising are:
(1) What variables affect the disease and in what way?
(2) Should surgical intervention focus on single or on multiple fingers?
The first is an epidemiological question, the second is a clinical one.

Empirical research has described the patterns of occurrence of Dupuytren disease in multiple fingers. 
\cite{meyerding1936dupuytren} have stated that most often the combination of affected ring and little fingers occurred, followed by the combination of an affected third, fourth, and fifth finger.
\cite{tubiana1982location} found that Dupuytren disease rarely affects radial side alone, and that the radial affect is often associate with an affected ulnar side.
\cite{milner2003dupuytren} noticed that patients who had required surgery because of an affected thumb were on average 8 years older 
and had suffered significantly longer from the disease compared to patients with a mildly affected radial side.
Moreover, these patients suffered from ulnar disease that repeatedly had required surgery, suggesting an intractable form of disease.
More recently, \cite{lanting2014patterns}, with a multivariate ordinal logit model, suggested that the middle finger is substantially correlated with other fingers on the ulnar side, and the thumb and index finger are correlated.
They took into account age and sex, and tested hypotheses on independence between groups of fingers.
However, so far, no serious multivariate analysis of the disease has been performed taking into account potential risk factors.

Essential risk factors of Dupuytren disease include both phenotypic and genotypic factors \citep{shih2010scientific}, 
such as genetic predisposition and ethnicity, as well as sex and age.
However, it is unclear whether Dupuytren disease is a complex oligogenic or a simple monogenic Mendelian disorder.
Several life-style risk factors (some considered controversial) include smoking, excessive alcohol consumption, manual work, and hand trauma. 
Also several diseases, such as diabetes mellitus and epilepsy, are thought to affect the life-style factors risk of incurring Dupuytren disease.
However, the role of these life-style factors and diseases has not been fully elucidated, and the results of different studies are occasionally conflicting \citep{lanting2014systematic}.

In this paper we analyze data collected by the Department of Plastic Surgery of the University Medical Center Groningen 
in the north of Netherlands from patients who have Dupuytren disease.
The data are originally described by \cite{lanting2013prevalence} and \cite{lanting2014patterns}.
Both hands of patients are examined for signs of Dupuytren disease.
These are tethering of the skin, nodules, cords, and finger contractures in patients with cords. 
Severity of the disease is measured by the angles on each of the 10 fingers.
Recorded potential risk factors include smoking habits, alcohol consumption, whether participants had performed manual labor during a significant part of their life, 
and whether they had sustained hand injury in the past, including surgery. 
In addition, information about the presence of Ledderhose diabetes, epilepsy, peyronie, knuckle pad, liver disease, 
and familial occurrence of Dupuytren disease, defined as a first-degree relative with Dupuytren disease, was collected.
% The data consisted of $279$ patients with Dupuytren disease.
% In $79$ of them, at least one of their fingers have been affected by the disease and show signs of contraction.

The primary aim of this paper is to model the relationships between the risk factors and disease indicators for Dupuytren disease based on the mixed dataset.
We propose a computationally efficient Bayesian statistical method based on Gaussian copula graphical models (GCGMs)
for discovering the joint conditional independence structure of binary, ordinal or continuous variables simultaneously.
Our Bayesian framework is based on that proposed by \cite{dobra2011copula}.
In GCGMs, the graph selection procedure is embedded inside a semiparametric framework, using the extended rank likelihood \citep{hoff2007extending}.
In this paper we design our proposed Bayesian framework for GCGMs based on a computationally efficient search algorithm,
using a trans-dimensional MCMC approach based on a continuous-time birth-death process \citep{mohammadi2015bayesianStructure}.

In this work, we follow a similar approach as that of \cite{mohammadi2015bayesianStructure} by using their birth-death MCMC algorithm. 
Changes have been made to allow more general data structures of mixed type using copula graphical models.
Importantly we have made substantial improvements in their algorithm to overcome the computational bottle-neck in evaluating normalizing constants. 
Furthermore, our proposed approach can handle missing data without any additional computational effort, if the missingness is completely at random (MCAR).

In Section \ref{sec:methodology}, we illustrate our Bayesian framework based on Gaussian copula graphical models.
In addition, we show the performance of our method and we compare it to state-of-the-art alternatives.
In section \ref{sec: dupuytren data}, we analyze the Dupuytren disease dataset based on the proposed Bayesian birth-death MCMC method.
In this section, first analyze potential phenotype risk factors for Dupuytren disease.
Moreover, we analyze the relationship between consider the severity of Dupuytren disease of pairs of fingers.
The result helps surgeons to decide whether they should operate one finger or they should operate multiple fingers simultaneously. 
Finally, we discuss the connections between existing methods and possible future directions. % in the last section.

%%%%%%%%%%%%%%%%%%%%%%%%%%%%%%%%%%%%%%%%%%%%%%%%%%%%%%%%%%%%%%%%%%%%%%%%%%%%%%%%%%%%%%%%%%%%%%%%%%%%%%%%%%%%%%%%%%%%%%%%%%%%%%%%%%%%%%%%%%%%%%%%%%%%%%%%%%%%%%%%%%%%%
\section{Methodology}
\label{sec:methodology}

% %%%%%%%%%%%%%%%%%%%%%%%%%%%%%%%%%%%%%%%%%%%%%%%%%%%%%%%%%%%%%%%%%%%%%%%%%%%%%%%%%%%%%%%%%%%%%%%%%%%%%%%%%%%%%%%%%%%%%%%%%%%%%%%%%%%%%%%%%%%%%%%%%%%%%%%%%%%%%%%%%%%%%
% \subsection{Bayesian graphical models}
% \label{subsec:Bayesian GMs}

Graphical models \citep{lauritzen1996graphical} provide an effective way to describe statistical patterns in multivariate data. %, specially for high-dimensional datasets such as gene expression data. 
In this context undirected Gaussian graphical models are commonly used, since inference in such models is often tractable.
In undirected Gaussian graphical models, the graph structure is characterized by its precision matrix (the inverse of covariance matrix): 
the non-zero entries in the precision matrix show the edges in the conditional independence graph. 
In the real world, data are often non-Gaussian, like the dataset considered in our application.
For non-Gaussian continuous data, variables can be transformed to Gaussian latent variables.
For discrete data, however, there is no one-to-one transformation into latent Gaussian variables.
A common approach is to apply a Markov chain Monte Carlo method (MCMC) to simulate both the latent Gaussian variables and the posterior distributions \citep{hoff2007extending}.
Another Bayesian approach is the Gaussian copula graphical models developed by \cite{dobra2011copula},
in which the sampler algorithm is based on reversible-jump MCMC and a Cholesky decomposition of the precision matrix. 
Alternatively, our proposed method implements the birth-death MCMC approach \citep{mohammadi2015bayesianStructure} which has several computational advantages
compared to reversible-jump MCMC approach as we show in our simulation examples. %; see \cite[Section 4]{mohammadi2015bayesianStructure}.

For studying the multivariate dependence structure of finger contractures and risk factors, 
we are interested in exploring the conditional independence graph space and identifying conditional independence graphs that are most appropriate for our given data. 
In this regard, we calculate the posterior distribution of the conditional independence graph $G$ conditional on data
\[
 Pr( G | \mbox{data} ) = \frac{Pr(G) Pr(\mbox{data}|G)} {\sum_{G \in \mathcal{G}} Pr(G) Pr(\mbox{data}|G)},
\]
in which $\mathcal{G}$ is the conditional independence graph space. 
Computing this posterior distribution is computationally unfeasible, since in the denominator we require the sum over all possible graphs. 
The graph space increases super-exponentially with the dimension of the variables.
For $p$ nodes in a conditional independence graph, there are ${p(p-1)/2}$ possible edges, and hence we have $2^{p(p-1)/2}$ different possible graphs
corresponding to all combinations of individual edges being in or out of the graph.
For example, in our data we have $23$ variables ($p=23$), resulting in a total number of possible conditional independence graphs of more than $10^{76}$.
This motivates us to develop effective search algorithms for exploring graphical model uncertainty. 
In order to be accurate and scalable, the key is to design computationally efficient search algorithms that are able to quickly move towards high posterior probability regions, 
and to take advantage of local computations. 

%%%%%%%%%%%%%%%%%%%%%%%%%%%%%%%%%%%%%%%%%%%%%%%%%%%%%%%%%%%%%%%%%%%%%%%%%%%%%%%%%%%%%%%%%%%%%%%%%%%%%%%%%%%%%%%%%%%%%%%%%%%%%%%%%%%%%%%%%%%%%%%%%%%%%%%%%%%%%%%%%%%%%
\subsection{Gaussian copula graphical models}
\label{subsec:CGGM}

In graphical models, conditional dependence relationships among random variables are presented as a graph $G$.
A graph $G=(V,E)$ specifies a set of vertices $V = \{1, 2, . . . , p\}$, where each vertex corresponds to a random variable, and a set of edges $E \subset V\times V$ \citep{lauritzen1996graphical}.
$\Ehat$ denotes the set of non-existing edges.
We focus here on undirected graphical models, where $(i,j) \in E \Leftrightarrow (j,i) \in E$, also known as Markov random fields. 
The absence of an edge between two vertices specifies the pairwise conditional independence of these two variables given the remaining variables,
while an edge between two variables determines the conditional dependence of the variables.
In our application, for example, disease risk factors (such as disease factors, alcohol, and hand injury) will be the nodes, and dependencies among them will be the edges.

A graphical model that follows a multivariate normal distribution is called a Gaussian graphical models, also known as a covariance selection model \citep{dempster1972covariance}.
Zero entries in the precision matrix correspond to the absence of edges on the graph and conditional independence between pairs of random variables
given all other variables. 
We define a zero mean Gaussian graphical model with respect to the graph $G$ as
\begin{eqnarray*}
\mathcal{M}_G = \left\{ \mathcal{N}_p (0,\Sigma) \ | \ K=\Sigma^{-1} \in \mathbb{P}_{G} \right\},
\end{eqnarray*}
where $\mathbb{P}_{G}$ denotes the space of $p\times p$ positive definite matrices with entries $(i,j)$ equal to zero whenever $(i,j) \in \Ehat$. 
Let $ \mathbf{z} = (\mathbf{z}^{1},...,\mathbf{z}^{n}) $ be an independent and identically distributed sample of size $n$ from model $\mathcal{M}_G$, where $\mathbf{z}^{i}$ is a $p$ dimensional vector of variables. 
Then, the likelihood is
\begin{eqnarray}
\label{likelihood}
P( \mathbf{z} | K,G) \propto |K|^{n/2} \exp \left\{ -\frac{1}{2} \mbox{tr}(KS) \right\},
\end{eqnarray}
where $S = \mathbf{z}' \mathbf{z}$.

In line with extending the idea of Gaussian graphical modeling for non-Gaussian data, the Gaussian
copula has been considered for graphical modeling among a set of mixed variables \citep{dobra2011copula}. 
A copula is a multivariate cumulative distribution function whose uniform marginals are on the interval $[0,1]$. 
Copulas provide a flexible tool for understanding dependence among random variables, in particular for non-Gaussian multivariate data, for example, 
the type of data application we consider in this study which consists of  $23$ mixed variables ($13$ phenotype risk factors and $10$ variables on the severity of 
Dupuytren disease in all the hand fingers) of the type discrete, binary, and ordered categorical variables; see Section \ref{subsec: risk factors}.

By Sklar's theorem \citep{sklar1959fonctions} there exists a copula $C$ such that any $p$ dimensional distribution function $H$ can be completely 
specified by its marginal distributions $F_j, ~~j=1, \ldots, p$, and a copula $C$ satisfying
\begin{equation*}
 H(y_1, \ldots, y_p) = C \left(F_1(y_1), \ldots, F_p(y_{p})\right).
\end{equation*}
Conversely, a copula function can be extracted from any $p$ dimension distribution function $H$ and marginal distributions $F_j$ by 
\begin{equation*}
 C(u_1, \ldots, u_p ) = H \left(F^{-1}_1(y_1), \ldots, F^{-1}_p(y_{p})\right) ,
\end{equation*}
where $F^{-1}_j (s) = \inf \{t \mid F_j(t) \geq s \}$ are the pseudo-inverse of $F_j$. The Gaussian copula is given by
\[
 C(u_1, \ldots, u_p \mid \Gamma) = \Phi_p\left(\Phi^{-1}(u_1), \ldots, \Phi^{-1}(u_p) \mid \Theta\right),
\]
where $u_j = F_j(y_j), ~j=1, \ldots, p$, $\Phi_p(\cdot)$ is the cumulative distribution of a multivariate normal distribution and $\Phi(\cdot)$ is a cumulative distribution 
function of a univariate normal distribution. %It follows that $Y_j = F_j^{-1}\left(\Phi(Z_j)\right)$.

The decomposition of a joint distribution into marginal distributions and a copula suggests that the copula captures the essential dependence features between random variables. 
Moreover, the copula measure of dependence is invariant to any monotone transformation of the random variables. 
Thus, copulas allow one to model the marginal distributions and the dependence structure of a multivariate random variables separately. 
In copula modeling, \citet{genest1995semiparametric} develop a popular semiparametric estimation approach or rank likelihood based estimation 
in which the association among variables is represented with a parametric copula but the marginals are treated as nuisance parameters. 
The marginals are estimated non-parametrically using the scaled empirical distribution function $\hat{F}_j(y) = \frac{n}{n+1} F_{n_j} (y)$, 
where $F_{n_j} (y) = \frac{1}{n}\sum^n_{i=1} I\{y_{ij} \leq y\}$. 
As a result estimation and inference are robust to misspecification of marginal distributions. 

However, for discrete data, where the distribution of ranks depends on the 
univariate marginal distributions, the use of the semiparametric estimator is somewhat inappropriate for the analysis of mixed continuous and discrete data \citep{hoff2007extending}.   
To overcome this, he proposes the extended rank likelihood, which is a type of marginal likelihood 
approach where the ranks are free of the marginal distributions of the discrete data. 
This makes the extended rank likelihood approach more appropriate for graphical modeling that 
avoids the difficult problem of modeling the marginal distributions. 

Let $Y$ be a collection of continuous, binary, ordinal or count variables with $F_j$ the marginal distribution of $Y_j$ and $F_j^{-1}$ its pseudo inverse. 
For constructing a joint distribution of $Y$, we introduce a multivariate normal latent variable as follows
\begin{eqnarray*}
Z_1, ..., Z_n \stackrel{iid}{\sim} \mathcal{N}(0, \Sigma),
\end{eqnarray*}
where $\Sigma$ is the correlation matrix.
We define the observed data as
\begin{eqnarray*}
Y_{ij} = F_j^{-1}(\Phi(Z_{ij})).
\end{eqnarray*}
A Gaussian copula-based joint cumulative distribution of $\mathbf{Y}$ is given by
\begin{eqnarray}
\label{jointdf}
 P \! \left(Y_1 \! \leq \! y_1, \ldots, Y_p \! \leq \! y_p \right) \! = \! \Phi_p \! \left( \! \Phi^{-1} \! (F_1(y_1)), \ldots, \Phi^{-1} \! (F(y_p)) \! \mid \! \Gamma\right) \!.
\end{eqnarray}

Our aim is to infer the underlying graph structure $G$ of the mixed variables $\mathbf{Y}$ implied by the continuous latent variables $\mathbf{Z}$.  
The observed mixed data from a sample of $n$ observations $\mathbf{y}$ are related to the latent samples $\mathbf{z}$ that belong to the set 
\begin{eqnarray}
  \mathcal{A}(\mathbf{y}) &=& \left\{\mathbf{z} \in \mathbb{R}^{n \times p} : \max\left\{ z_j^{(k)}: y_j^{(s)} < y_j^{(r)} \right\} < z_j^{(r)} < \min\left\{ z_j^{(s)}: y_j^{(r)} < y_j^{(s)}\right\} \right. , \nonumber\\
&& ~~~~ \left. r,k,s=1, \ldots, n; j=1, \ldots, p \right\}  \label{truncated set}. 
\end{eqnarray}

It follows that inference on the latent space can be performed by substituting the observed data $\mathbf{y}$ with 
the event $\mathbf{z} \in \mathcal{A}(\mathbf{y})$. 
For a given graph $G$ and precision matrix $K = \Sigma^{-1}$, the extended rank likelihood is defined as
\begin{eqnarray}
 P \! (\mathbf{z} \in \mathcal{A}(\mathbf{y}) \! \mid \! K, G) \! = \! P(\mathbf{z} \in \mathcal{A}(\mathbf{y}) \mid K, G) \! = \! \int_{\mathcal{A}(\mathbf{y})} \! P(\mathbf{z} \! \mid \! K, G) d \mathbf{z}, \label{exranklik}
\end{eqnarray}
where the expression inside the integral for the Gaussian copula based probability function given by \eqref{jointdf} takes a similar form as in \ref{likelihood}. 
In the next sections we develop a Bayesian approach based on the extended rank likelihood given in \eqref{exranklik}.

%%%%%%%%%%%%%%%%%%%%%%%%%%%%%%%%%%%%%%%%%%%%%%%%%%%%%%%%%%%%%%%%%%%%%%%%%%%%%%%%%%%%%%%%%%%%%%%%%%%%%%%%%%%%%%%%%%%%%%%%%%%%%%%%%%%%%%%%%%%%%%%%%%%%%%%%%%%%%%%%%%%%%
\subsection{Bayesian Gaussian copula graphical models}
\label{subsec:Bayesian CGGM}

In the Bayesian framework that follows, we infer about graph and precision matrix $(K,G)$ by considering a posterior distribution 
\begin{eqnarray}
\label{joint posterior}
P(K,G | \mathbf{z} \in \mathcal{A}(\mathbf{y})) \propto P(\mathbf{z} \in \mathcal{A}(\mathbf{y}) | K, G) P(K \mid G) P(G), 
\end{eqnarray}
% which is discussed in detail in the next sections.
where $P(\mathbf{z} \in \mathcal{A}(\mathbf{y}) | K, G)$ is the extended rank likelihood defined in \eqref{exranklik}, 
$P(K \mid G)$ denotes a prior distribution of a precision matrix $K$ for a given graph structure and $P(G)$ denotes a prior distribution for a graph $G$.

In particular, we develop a simple and efficient continuous birth-death Markov chain Monte Carlo (MCMC)
algorithm for the posterior computation that converges much faster than the reversible MCMC algorithm in \cite{dobra2011copula} and overcomes the computational bottle-neck in \cite{mohammadi2015bayesianStructure}.  
Moreover, we evaluate the results induced by the latent variables using posterior predictive analysis on the scale of the original mixed variables. 

%%%%%%%%%%%%%%%%%%%%%%%%%%%%%%%%%%%%%%%%%%%%%%%%%%%%%%%%%%%%%%%%%%%%%%%%%%%%%%%%%%%%%%%%%%%%%%%%%%%%%%%%%%%%%%%%%%%%%%%%%%%%%%%%%%%%%%%%%%%%%%%%%%%%%%%%%%%%%%%%%%%%%
\subsubsection{Prior specification}
\label{subsec:priors}

In what follows we briefly describe the specification of prior distributions for the graph $G$ and the precision matrix $K$.
For the prior distribution of the graph, we propose to use a discrete uniform distribution over the graph space, as a non-informative prior. 
We also note that other choices of priors for the graph structure have been considered by modeling the joint state of the edges \citep{scutari2013prior}, 
encouraging sparse graphs \citep{jones2005experiments} or a truncated Poisson distribution on the graph size \citep{mohammadi2015bayesianStructure}.

We consider the G-Wishart \citep{roverato2002hyper} distribution as prior distribution of the precision matrix.
The G-Wishart is the Wishart distribution restricted to the space of precision matrices with zero entries specified by a graph $G$, $\mathbb{P}_G$.
The G-Wishart density for $K \in \mathbb{P}_G \sim W_G(b,D)$ can be written as
$$
P(K|G)=\frac{1}{I_G (b,D)} |K|^{(b-2)/2} \exp \left\{ -\frac{1}{2} \mbox{tr}(DK) \right\},
$$
where $b > 2$ is the degree of freedom, $D$ is a symmetric positive definite matrix, and $I_G (b,D)$ is the normalizing constant,
$$
I_G (b,D) = \int_{\mathbb{P}_{G}} |K|^{(b-2)/2} \exp \left\{ -\frac{1}{2} \mbox{tr}(DK) \right\} dK.
$$
Dealing with this normalizing constant has been a major issue in recent literature; see Section \ref{paragraph:computing ratio}.

Since the G-Wishart prior is conjugate to normally distributed data \eqref{likelihood}, the posterior distribution of $K$ condition on graph $G$ is
\[
P(K|Z \in \mathcal{A}(\mathbf{y}) ,G) = \frac{1}{I_G (b^*,D^*)} |K|^{(b^*-2)/2} \exp \left\{ -\frac{1}{2} \mbox{tr}(D^*K) \right\},
\]
where  $b^*=b+n$ and $D^*=D+S$ with $S = \mathbf{z}' \mathbf{z}$, that is a G-Wishart distribution, $W_G(b^*,D^*)$.
For other choices of priors for the precision matrix see \cite{wang2013class, wang2015scaling, wang2012bayesian, wong2003efficient}.

%%%%%%%%%%%%%%%%%%%%%%%%%%%%%%%%%%%%%%%%%%%%%%%%%%%%%%%%%%%%%%%%%%%%%%%%%%%%%%%%%%%%%%%%%%%%%%%%%%%%%%%%%%%%%%%%%%%%%%%%%%%%%%%%%%%%%%%%%%%%%%%%%%%%%%%%%%%%%%%%%%%%%
\subsubsection{Sampling algorithm for posterior inference}
\label{subsubsec:posterior inference}

Sampling from the joint posterior distribution \eqref{joint posterior} can be done by a computationally efficient birth-death MCMC algorithm
proposed in \cite{mohammadi2015bayesianStructure} for Gaussian graphical models.
Here we extend their algorithm for the more general case of Gaussian copula graphical models.
Our algorithm is based on a continuous time birth-death Markov process in which the algorithm explores the graph space by adding or removing an edge in a birth or death event, respectively.
The birth and death rates of edges are determined by the stationary distribution of the process.
The algorithm is designed in such a way that the stationary distribution equals to the target joint posterior distribution of the graph and the precision matrix \eqref{joint posterior}. 
The time between two successive events has an exponential distribution. %with mean $1/(\beta(K)+\delta(K))$. 
Therefore, the probability of birth and death events are proportional to their rates.

\citet[section 3]{mohammadi2015bayesianStructure} proved that the birth-death MCMC (BDMCMC) algorithm converges to the target joint posterior distribution of the graph and the precision matrix, 
by considering the following birth and death rates,
\begin{equation}
\label{birthrate}
\beta_{e}(K) = \frac{P(G^{+e},K^{+e} \setminus(k_{ij},k_{jj})|Z \in \mathcal{A}(\mathbf{y}))}{P(G,K\setminus k_{jj}|Z \in \mathcal{A}(\mathbf{y}))}, \ \ \mbox{for each} \ \ e \in \Ehat,
\end{equation}
\begin{equation}
\label{deathrate}
\delta_{e}(K) =\frac{P(G^{-e},K^{-e}\setminus k_{jj}|Z \in \mathcal{A}(\mathbf{y}))}{P(G,K\setminus(k_{ij},k_{jj})|Z \in \mathcal{A}(\mathbf{y}))}, \: \: \qquad \mbox{for each} \ \ e \in E,
\end{equation}
in which $G^{+e}=(V, E \cup \{ e \})$ for the inclusion of an edge from the graph $G$, and $K^{+e} \in \mathbb{P}_{G^{+e}}$ represents an updated precision matrix 
when an edge is included, and similarly $G^{-e} = ( V, E \setminus \{ e \})$, and $K^{-e} \in \mathbb{P}_{G^{-e}}$ represent exclusion of an edge and its updated precision matrix.
We refer our proposed approach as the extended birth-death MCMC (EBDMCMC) algorithm. Details of the EBDMCMC algorithm are given in Algorithm 1.  
\begin{algorithm}
% \label{copulabdmcmc}
{\bf Algorithm 1} Given a graph $G=(V,E)$ with a precision matrix $K$, iterate the following steps:
\begin{description}
 \item[1.] Sample the latent data. For each $r \in V$ and $j \in \{ 1,2,...,n \}$, we update the latent value $z_r^{(j)}$ 
 from its full conditional distribution
 \begin{eqnarray}
 Z_r | K, Z_{V \setminus \{ r \} } = z_{K, V \setminus \{ r \} }^{(j)} \sim  \mathcal{N} \left(- \sum_{r'} { K_{rr'} z_{r'}^{(j)} / K_{rr} }, 1/K_{rr} \right), \label{tmvn}
 \end{eqnarray}
truncated to the interval 
in \eqref{truncated set}. This sampling step can be easily modified to handle data that are missing-at-random. 
That is if $y_r$ is missing, then the full conditional $Z_r | K, Z_{V \setminus \{ r \} } $ is the untruncated multivariate normal distribution given in \eqref{tmvn}.

 \item[2.] Sample the graph based on birth and death process.  
  \begin{description}
   \item[2.1.] Calculate the birth rates by equation \ref{birthrate} and $\beta(K)= \sum_{e \in \Ehat}{\beta_{e}(K)}$,
   \item[2.2.] Calculate the death rates by equation \ref{deathrate} and $\delta(K)=\sum_{e \in E}{\delta_{e}(K)}$,
   \item[2.3.] Calculate the waiting time by $W(K)= 1/(\beta(K)+\delta(K))$,
   \item[2.4.] Calculate the jump type (birth or death),
  \end{description}
 \item[3.] Sample the new precision matrix, according to the jump type, based on Algorithm 2.
\end{description}
\end{algorithm}

In Algorithm 1,  the first step is to sample the latent variables given the observed data.
Then, based on this sample, we calculate the birth and death rates and waiting times.
The birth and death rates are used to calculate the jump type. 
Details of how to efficiently calculate the birth and death rates are discussed in Section \ref{subsubsec: computing rates}. 
Finally in step 3, according to the jump, we sample a new precision matrix using a direct sampling scheme from the G-Wishart distribution which is described in Algorithm 
2 in Section \ref{subsubsec: sampling G-Wishart}.

\begin{figure} [!ht]
\centering
\includegraphics[width=0.9\textwidth]{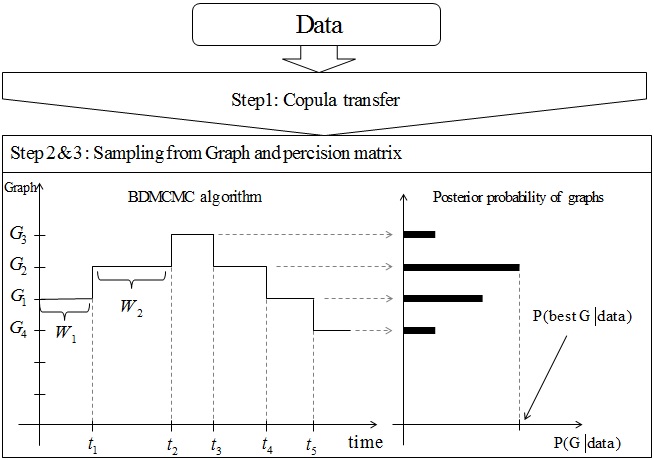}
\caption{ \label{fig:EBDMCMC} This image visualizes Algorithm 1.
  (On top) Mixed observed data transformation using the copula to sample the latent variables, (Bottom left) Continuous time EBDMCMC algorithm where $ \left\{ W_1, W_2,... \right\}$ 
  denote waiting times and $ \left\{ t_1, t_2,... \right\}$ denote jumping times. 
  (Bottom right) Estimated posterior probability of the graphs which are proportional to sum of their waiting times. }
\end{figure}

For our algorithm, the Rao-Blackwellized sample mean \citep[subsection 2.5]{cappe2003reversible} provides an effective way to estimate the posterior probability of each graph.
The Rao-Blackwellized estimate of the posterior graph probability is the proportion to the total waiting times for that graph (see Figure \ref{fig:EBDMCMC} in the right). 
The waiting times for each graph act as the weights of that graph (e.g. $ \left\{ W_1, W_2, W_3,... \right\}$ in Figure \ref{fig:EBDMCMC} in the left).

%%%%%%%%%%%%%%%%%%%%%%%%%%%%%%%%%%%%%%%%%%%%%%%%%%%%%%%%%%%%%%%%%%%%%%%%%%%%%%%%%%%%%%%%%%%%%%%%%%%%%%%%%%%%%%%%%%%%%%%%%%%%%%%%%%%%%%%%%%%%%%%%%%%%%%%%%%%%%%%%%%%%%
\subsubsection{Computing the birth and death rates}
\label{subsubsec: computing rates}

Calculating the birth and death rates \eqref{birthrate} and \eqref{deathrate} has been a major bottle-neck of the EBDMCMC algorithm. 
Here, we explain how to resolve the computational bottle-neck and come-up with an efficient way to calculate the death rates; the birth rates are calculated in a similar manner.

Following \cite{mohammadi2015bayesianStructure} and after some simplification, for each $e=(i,j) \in E$, we have
\begin{eqnarray}
\label{death rates}
\delta_{e} (K) = \frac{P(G^{-e})}{P(G)} \frac{I_{G}(b,D)}{I_{G^{-e}}(b,D)} (\frac{D^*_{jj}}{2 \pi(k_{ii}-k^1_{11})}) ^ {\frac{1}{2}} H(K,D^*),
\end{eqnarray}
where 
\begin{eqnarray*}
H(K,D^*) \! = \! \exp \! \left\{ \! - \! \frac{1}{2} \! \left[ \! \mbox{tr}(D_{e,e}^{*}(K^0-K^1)) - (D^*_{ii} - \frac{(D^*_{ij})^2}{D^*_{jj}}) (k_{ii}-k^1_{11}) \! \right] \! \right\},
\end{eqnarray*}
in which 
\begin{eqnarray*} % \footnotesize
K^0 =
\begin{bmatrix}
 k_{ii}  & 0                                                                                 \\
  0      & K_{j,V \setminus j} (K_{V \setminus j, V \setminus j})^{-1} K_{V \setminus j, j}  \\
\end{bmatrix},
\end{eqnarray*}
and $K^1 = K_{e,V \setminus e} (K_{V \setminus e, V \setminus e})^{-1} K_{V \setminus e, e}$.
The computational bottle-neck in \eqref{death rates} is the ratio of normalizing constants, $\frac{I_{G}(b,D)}{I_{G^{-e}}(b,D)}$.

%%%%%%%%%%%%%%%%%%%%%%%%%%%%%%%%%%%%%%%%%%%%%%%%%%%%%%%%%%%%%%%%%%%%%%%%%%%%%%%%%%%%%%%%%%%%%%%%%%%%%%%%%%%%%%%%%%%%%%%%%%%%%%%%%%%%%%%%%%%%%%%%%%%%%%%%%%%%%%%%%%%%%
\paragraph{Dealing with calculation of normalizing constants.}
\label{paragraph:computing ratio}

Calculating the ratio of normalizing constants has been a major issue in recent literature \citep{uhler2014exact, wang2012efficient, mohammadi2015bayesianStructure}. 
To compute the normalizing constants of a G-Wishart, \cite{roverato2002hyper} proposed an importance sampling algorithm, while \cite{atay2005monte} developed a Monte Carlo method.
These methods can be computationally expensive and numerically unstable \citep{jones2005experiments, wang2012efficient}.
\cite{wang2012efficient, cheng2012hierarchical, mohammadi2015bayesianStructure} developed an alternative approach, which borrows ideas from the exchange algorithm \citep{murray2012mcmc} and the 
double Metropolis-Hastings (MH) algorithm \citep{liang2010double} to compute the ratio of such normalizing constants. 
However, it has been observed that in case of high dimensional setting (large p), the double MH sampler become computationally inefficient due to the curse of dimensionality. 
To remedy this, more recently \citet[theorem 3.7]{uhler2014exact} derived an explicit representation of the normalizing constant ratio. 
\begin{theorem}[\citealt{uhler2014exact}]
\label{theorem: normalizing constant}
let $G=(V,E)$ be an undirected graph and $G^{-e}=(V,E^{-e})$ denotes the graph $G$ with one less edge $e$.
Then
\begin{equation*}
\frac{I_{G}(b,\mathbb{I}_p)}{I_{G^{-e}}(b,\mathbb{I}_p)} = 2 \sqrt{\pi} \frac{\Gamma((b+d+1)/2)}{\Gamma((b+d)/2)},
\end{equation*}
where $d$ denotes the number of triangles formed by the edge $e$ and two other edges in $G$
and $\mathbb{I}_p$ denotes an identity matrix with $p$ dimension.
\end{theorem}

Therefore, for the case of $D = \mathbb{I}_p$, we have a simplified expression for the death rates in \eqref{death rates}, which is given by
\begin{eqnarray*}
\delta_{e} (K) = \frac{P(G^{-e})}{P(G)} \frac{\Gamma((b+d+1)/2)}{\Gamma((b+d)/2)} (\frac{2 D^*_{jj}}{(k_{ii}-k^1_{ii})}) ^ {\frac{1}{2}} H(K,D^*).
\end{eqnarray*}

%%%%%%%%%%%%%%%%%%%%%%%%%%%%%%%%%%%%%%%%%%%%%%%%%%%%%%%%%%%%%%%%%%%%%%%%%%%%%%%%%%%%%%%%%%%%%%%%%%%%%%%%%%%%%%%%%%%%%%%%%%%%%%%%%%%%%%%%%%%%%%%%%%%%%%%%%%%%%%%%%%%%%
\subsubsection{Sampling from posterior distribution of precision matrix}
\label{subsubsec: sampling G-Wishart}

Several sampling methods from a G-Wishart have been proposed for sampling the precision matrix; for a review of existing methods see 
\cite{wang2012efficient, mitsakakis2011metropolis, wang2010simulation}. 
Here we use an exact sampler algorithm developed by \cite{lenkoski2013direct} after determining the graph structure and its precision matrix using 
Algorithm 1 and summarized the details in Algorithm 2.
\begin{algorithm}
%  \label{algorithm:K sampler}
{\bf Algorithm 2. Direct sampler from precision matrix} \citep{lenkoski2013direct}. 
Given a graph $G=(V,E)$ with precision matrix $K$ determined by the jump type (birth for inclusion of an edge and death for exclusion of an edge) from Algorithm 1: 
\begin{description}
\item[1.] Set $\Sigma = K^{-1}$,
\item[2.] Repeat for $j = 1, ..., p$, until convergence:
    \begin{description}
    \item[2.1] Let $N_j \subset V$ be the set of variables that connected to $j$ in $G$. \\
      Form $\Sigma_{N_j}$ and $K^{-1}_{N_j,j}$ and solve
      $$\hat{ \beta_{j}^{*} } = \Sigma^{-1}_{N_j} K^{-1}_{N_j, j},$$
    \item[2.2] Form $\hat{ \beta_{j} } \in R^{p-1}$ by plugging zeroes in those locations not connected to $j$ in $G$ and padding the elements of $\hat{ \beta_{j}^{*} }$ to the rest locations,
    \item[2.3] Replace $\Sigma_{j,-j}$ and $\Sigma_{-j,j}$ with $\Sigma_{-j,-j}\hat{ \beta_{j} }$,
    \end{description}
\item[3.] Return $K = \Sigma^{-1}$.
\end{description}
\end{algorithm}

%%%%%%%%%%%%%%%%%%%%%%%%%%%%%%%%%%%%%%%%%%%%%%%%%%%%%%%%%%%%%%%%%%%%%%%%%%%%%%%%%%%%%%%%%%%%%%%%%%%%%%%%%%%%%%%%%%%%%%%%%%%%%%%%%%%%%%%%%%%%%%%%%%%%%%%%%%%%%%%%%%%%%
\subsubsection{Simulation study}
\label{subsubsec:simulation study}

We perform a comprehensive simulation study with respect to different graph structures to evaluate the performance of the proposed EBDMCMC method and compare it to 
an alternative approach proposed by Dobra and Lenkoski \citep{dobra2011copula}, referred to as DL. 
We generate mixed data from a latent Gaussian copula model with 5 different types of variables, that includes 
\textquotedblleft Gaussian\textquotedblright, \textquotedblleft non-Gaussian\textquotedblright, \textquotedblleft ordinal\textquotedblright, 
\textquotedblleft count\textquotedblright, and \textquotedblleft binary\textquotedblright.
We performed all computations with our {\tt R} package {\tt BDgraph} \citep{bdgraph}. %, mohammadi2015BDgraph}.

\begin{figure} [!ht]
\centering
    \includegraphics[width=0.4\textwidth]{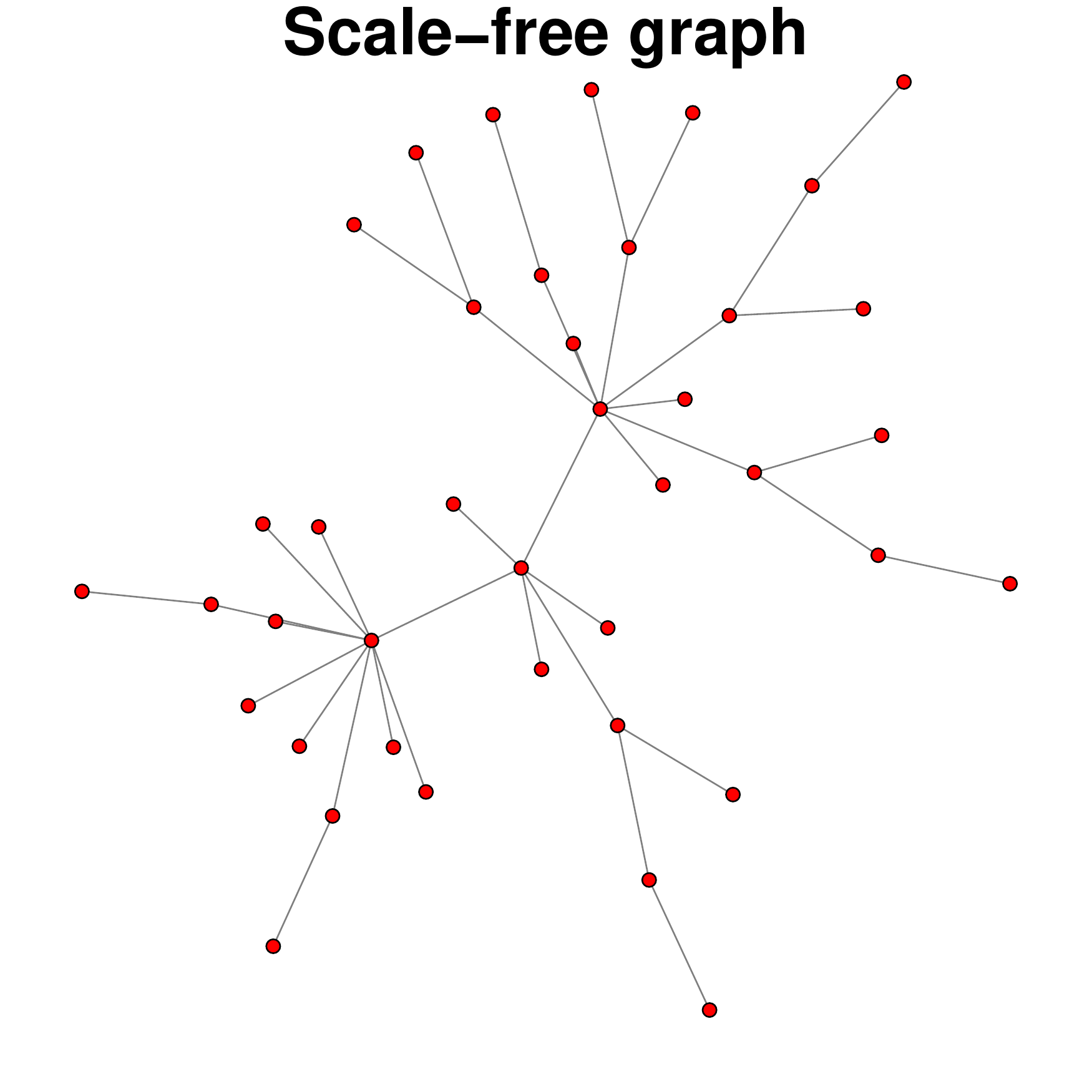}
    \includegraphics[width=0.4\textwidth]{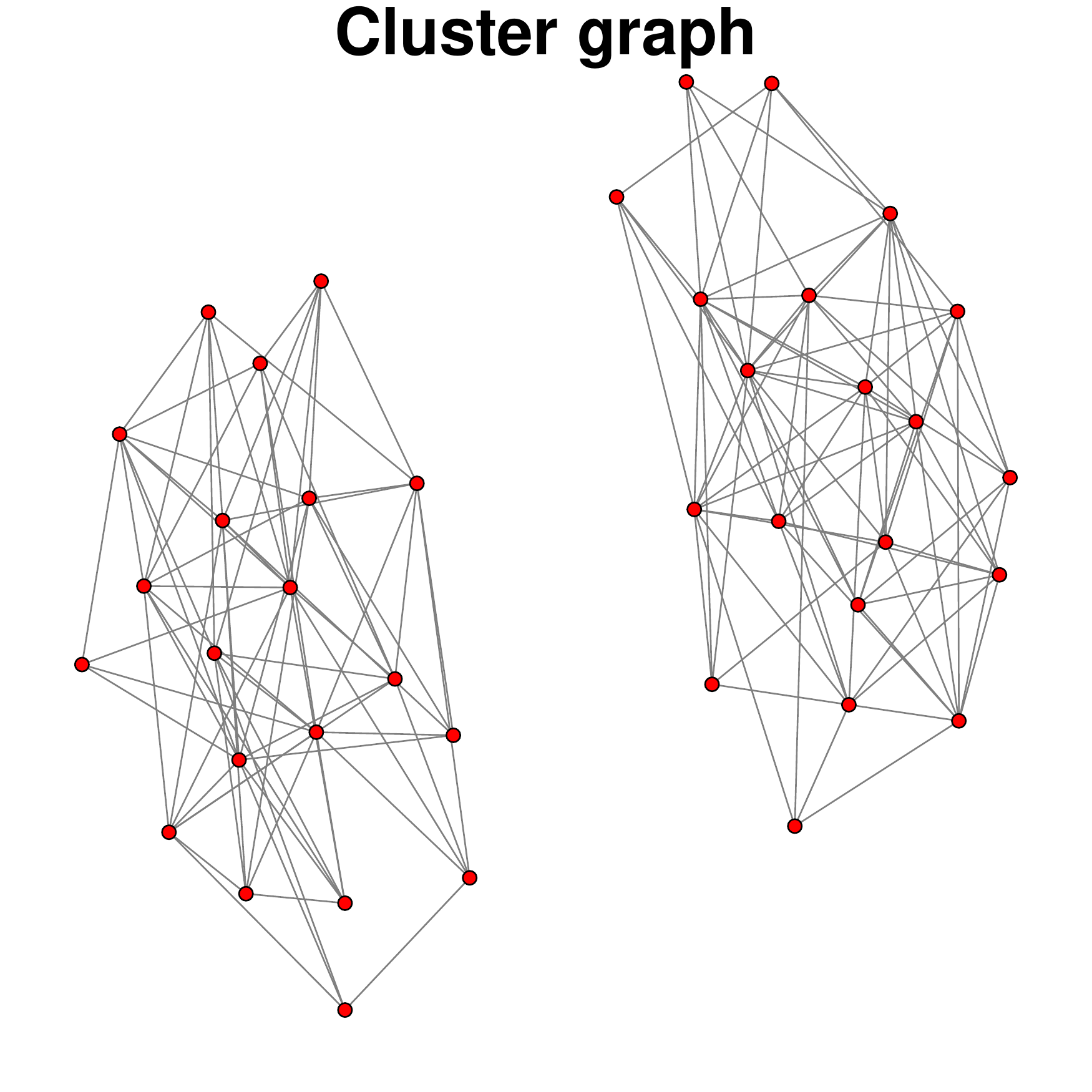}
    \includegraphics[width=0.4\textwidth]{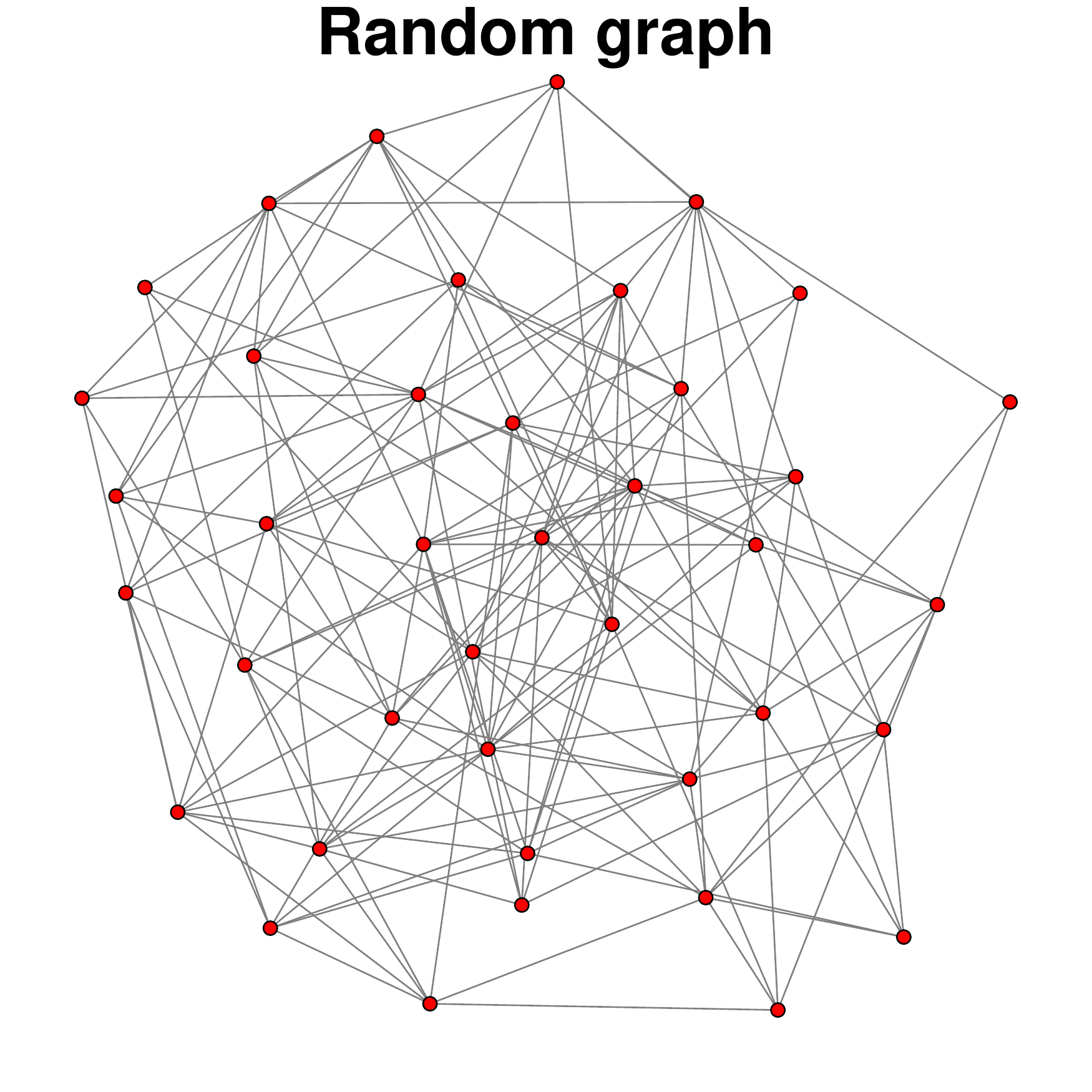}
\caption{ An illustration of the 3 simulated graph structures with $p=40$ nodes for our simulation example \ref{subsubsec:simulation study}. }
\label{fig:graphs}
\end{figure}

Corresponding to different sparsity patterns, we consider 3 different kinds of synthetic graphical models, having $p$ nodes:
\begin{itemize}
\item[1.] \textit{Random Graph:} A graph in which the edge set $E$ is randomly generated from independent Bernoulli distributions with 
                           probability $2/(p-1)$ and corresponding precision matrix is generated from $K \sim W_G(3,I_p)$.
\item[2.] \textit{Cluster Graph:} A graph in which the number of clusters is $ max \left\{ 2, \left[ p/20 \right] \right\} $. 
                            Each cluster has the same structure as a random graph.
                            The corresponding precision matrix is generated from $K \sim W_G(3,I_p)$.
\item[3.] \textit{Scale-free Graph:} A scale-free graph has a power-low degree distribution generated by the Barabasi-Albert algorithm \citep{albert2002statistical}.
                               The corresponding precision matrix is generated from $K \sim W_G(3,I_p)$.
\end{itemize}
Figure \ref{fig:graphs} presents the patterns for the 3 graph structures with $p=40$ nodes. 
For each graphical model, we consider different scenarios based on different number of variables $p=\{10,20,30,40 \}$ and different sample size; see Table \ref{table:F1-MSE}.
For each scenario, we generate mixed data and fit our proposed EBDMCMC 
and DL approaches using a uniform prior for the graph, $G$, and the G-Wishart prior $W_G(3, I_p)$ for the precision matrix, $K$. 
We run the two algorithms with the same starting points using $100,000$ iterations and $50,000$ iterations as a burn-in.
Computations for these scenarios were performed in parallel on a $235$ batch nodes with $12$ cores and $24$ GB of memory, running Linux.

To assess the performance of the graph structure, we compute the $F_1$-score measure \citep{powers2011evaluation} which is defined as
\begin{eqnarray}
\label{f1}
F_1\mbox{-score} = \frac{2 \mbox{TP}}{2 \mbox{TP + FP + FN}},
\end{eqnarray}
where TP, FP, and FN are the number of true positives, false positives, and false negatives, respectively. 
The $F_1$-score lies between $0$ and $1$, where $1$ stands for perfect identification and $0$ for bad identification.
Also, we use the mean square error (MSE), defined as
\begin{eqnarray}
\label{mse}
\mbox{MSE} =  \sum_{e}{ ( \hat{p}_e - I(e \in G_{true}) ) ^ 2 },
\end{eqnarray}
where $\hat{p}_e$ is the posterior edge inclusion probabilities and $I(e \in G_{true})$ is an indicator function, such that $I(e \in G_{true}) = 1$ if $e \in G_{true}$ and zero otherwise.
We calculate the posterior edge inclusion probabilities based on the Rao-Blackwellization \citep[subsection 2.5]{cappe2003reversible} for each possible edge $e = (i,j)$  as
\begin{eqnarray}
\label{posterior-edge}
\hat{p}_{e}= \frac{\sum_{t=1}^{N}{I(e \in G^{(t)}) W(K^{(t)}) }}{\sum_{t=1}^{N}{W(K^{(t)})}}, 
\end{eqnarray}
where $N$ is the number of iterations and $W(K^{(t)})$ is the waiting time for the graph $G^{(t)}$ with the precision matrix $K^{(t)}$. %; see \cite{mohammadi2015bayesianStructure}. 

\begin{table*}  [!ht] 
\caption{
Summary of performance measures in simulation example \ref{subsubsec:simulation study} for our method and DL \citep{dobra2011copula}. 
The table presents the $F_1$-score, defined in \eqref{f1} and MSE, defined in \eqref{mse}, with $50$ replications and standard deviations in parenthesis. 
The $F_1$-score reaches its best score at 1 and its worst at 0.
The MSE is positive value for which $0$ is minimal and smaller is better.
The best models for both $F_1$-score and MSE are boldfaced.
}
\label{table:F1-MSE} 
\begin{tabular}{l*{8}{l}l}
\hline
                   &                    &             & \multicolumn{2}{c}{F1-score}                    & \multicolumn{3}{c}{MSE}                             \\
                                                      \cmidrule{4-5}                                    \cmidrule{7-8}      %  \addlinespace                  \\
p                  & n                  & graph       & EBDMCMC                 & DL                     && EBDMCMC                  & DL                       \\
\hline                                                                                                                 %                                      \\
\multirow{4}{*}{10}&\multirow{4}{*}{30} & Random      &\textbf{0.52} (0.16)    & 0.38 (0.15)            && 6.44 (2.35)             &\textbf{6.33} (1.8)       \\
                   &                    & Cluster     &\textbf{0.58} (0.14)    & 0.43 (0.14)            &&\textbf{5.12} (1.75)     & 5.38 (1.3)               \\
                   &                    & Scale-free  &\textbf{0.53} (0.18)    & 0.43 (0.13)            && 6.46 (2.03)             &\textbf{6.43} (1.31)      \\
                   &                    &             &                        &                        &&                         &                          \\
\multirow{4}{*}{10}&\multirow{4}{*}{100}& random      &\textbf{0.71} (0.15)    & 0.67 (0.14)            &&\textbf{3.96} (1.73)     & 4.1 (1.42)               \\
                   &                    & Cluster     &\textbf{0.68} (0.16)    & 0.67 (0.16)            && 3.84 (1.54)             &\textbf{3.49} (1.14)      \\
                   &                    & Scale-free  &\textbf{0.67} (0.14)    & 0.63 (0.14)            && 4.5 (2)                 &\textbf{4.16} (1.18)      \\
                   &                    &             &                        &                        &&                         &                          \\
\multirow{4}{*}{20}&\multirow{4}{*}{70} & random      &\textbf{0.55} (0.08)    & 0.45 (0.06)            &&\textbf{23.28} (5.28)    & 24.04 (4.00)             \\
                   &                    & Cluster     &\textbf{0.55} (0.09)    & 0.47 (0.06)            && 23.67 (5.36)            &\textbf{21.84} (3.22)     \\
                   &                    & Scale-free  &\textbf{0.49} (0.13)    & 0.39 (0.08)            && 19.97 (6.29)            &\textbf{14.22} (1.96)     \\
                   &                    &             &                        &                        &&                         &                          \\
\multirow{4}{*}{20}&\multirow{4}{*}{200}& random      &\textbf{0.74} (0.07)    & 0.61 (0.07)            &&\textbf{15.10} (3.63)    & 17.27 (4.03)             \\
                   &                    & Cluster     &\textbf{0.73} (0.07)    & 0.62 (0.07)            &&\textbf{14.07} (4.00)    & 14.58 (3.87)             \\
                   &                    & Scale-free  &\textbf{0.67} (0.10)    & 0.55 (0.08)            && 11.3 (3.68)             &\textbf{9.38} (1.84)      \\
                   &                    &             &                        &                        &&                         &                          \\
\multirow{4}{*}{30}&\multirow{4}{*}{100}& Random      &\textbf{0.54} (0.06)    & 0.44 (0.04)            &&\textbf{52.3} (9.9)      & 59.1 (8.7)               \\
                   &                    & Cluster     &\textbf{0.56} (0.05)    & 0.47 (0.04)            &&\textbf{48.0} (6.5)      & 54.4 (8.1)               \\
                   &                    & Scale-free  &\textbf{0.53} (0.17)    & 0.30 (0.05)            && 27.7 (14.6)             &\textbf{25.8} (1.7)       \\
                   &                    &             &                        &                        &&                         &                          \\
\multirow{4}{*}{30}&\multirow{4}{*}{500}& Random      &\textbf{0.79} (0.04)    & 0.63 (0.07)            &&\textbf{25.8} (6.5)      & 41.1 (14.3)              \\
                   &                    & Cluster     &\textbf{0.79} (0.05)    & 0.66 (0.05)            &&\textbf{26.3} (5.2)      & 35.1 (7.9)               \\
                   &                    & Scale-free  &\textbf{0.81} (0.07)    & 0.59 (0.06)            &&\textbf{9.4} (3.2)       & 11.7 (3.0)               \\
                   &                    &             &                        &                        &&                         &                          \\
\multirow{4}{*}{40}&\multirow{4}{*}{400}& Random      &\textbf{0.71} (0.03)    & 0.57 (0.04)            &&\textbf{61.84} (7.77)    & 81.77 (14.25)            \\
                   &                    & Cluster     &\textbf{0.71} (0.04)    & 0.59 (0.04)            &&\textbf{58.25} (6.63)    & 69.4 (13.31)             \\
                   &                    & Scale-free  &\textbf{0.67} (0.09)    & 0.46 (0.06)            &&\textbf{23.26} (8.37)    & 23.29 (6.13)             \\
                   &                    &             &                        &                        &&                         &                          \\
\multirow{4}{*}{40}&\multirow{4}{*}{800}& Random      &\textbf{0.77} (0.03)    & 0.62 (0.06)            &&\textbf{50.07} (8.39)    & 76.22 (20.43)            \\
                   &                    & Cluster     &\textbf{0.80} (0.03)    & 0.68 (0.03)            &&\textbf{43.98} (5.93)    & 58.95 (7.27)             \\
                   &                    & Scale-free  &\textbf{0.78} (0.07)    & 0.55 (0.05)            &&\textbf{15.05} (5.14)    & 17.22 (5.4)              \\
\hline
\end{tabular}
\end{table*}
Table \ref{table:F1-MSE} reports comparisons of the EBDMCMC method with DL, where we repeat the experiments 50 times and report the average $F_1$-score and 
MSE with their standard errors in parentheses. 
Our EBDMCMC method performs well overall as its $F_1$-score is larger and its MSE is lower %are better 
than the DL method in most of the cases, mainly because of its fast convergence rate.
As we expected, the DL approach converges more slowly compared to our method.
% From a theoretical point of view, both algorithms converge to the true posterior distribution, if we run them a sufficient amount of time.
% Thus, the results from this table just indicate how quickly the algorithms converge. 

\begin{figure} [!ht]
\centering
    \includegraphics[width=0.75\textwidth]{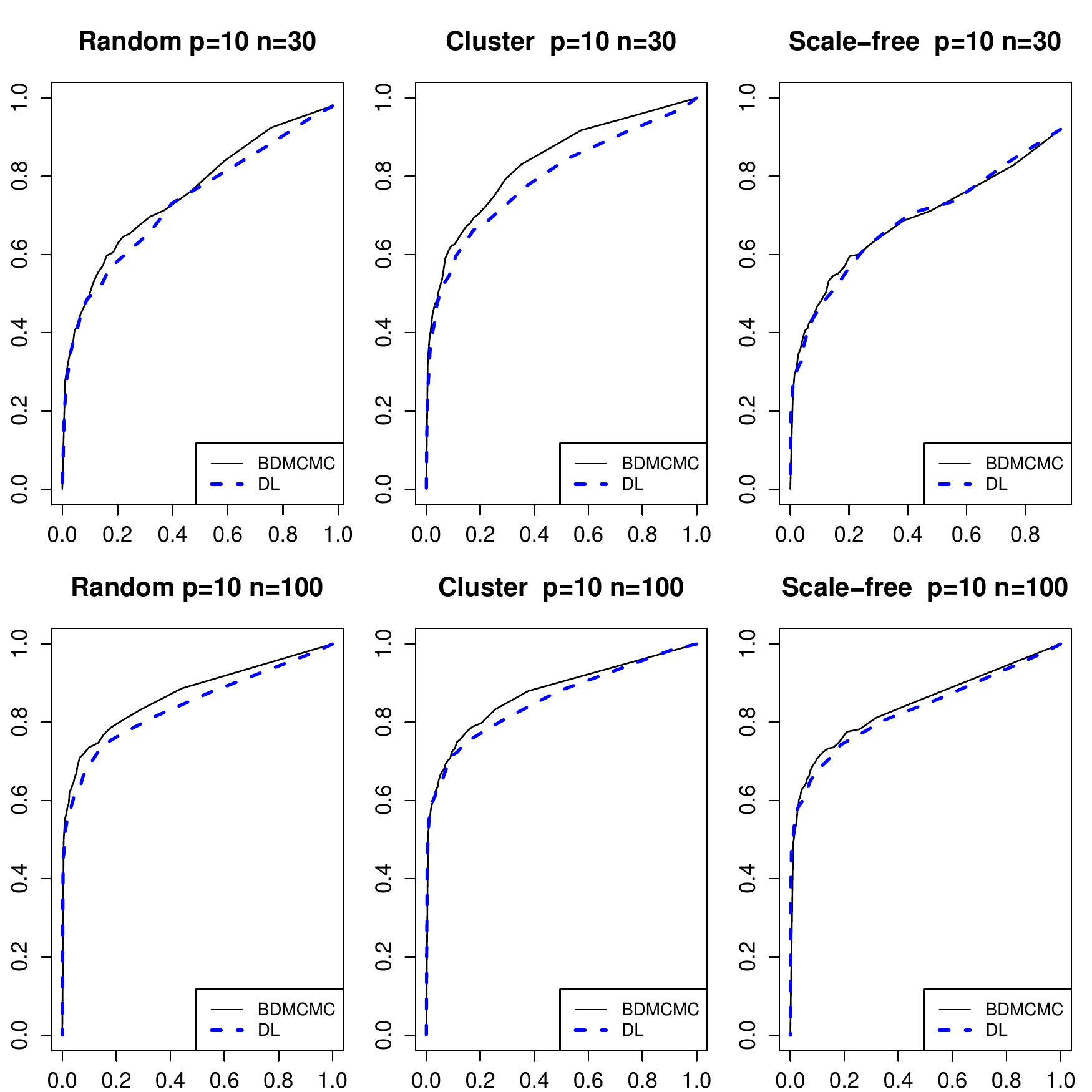}
\caption{ ROC curves depicting the performances of the simulated graphs for the case $p=10$ and $n=\{30,100\}$, based on the posterior edge inclusion probabilities. }
\label{fig:plotroc_p10}
\end{figure}

\begin{figure} [!ht]
\centering
    \includegraphics[width=0.75\textwidth]{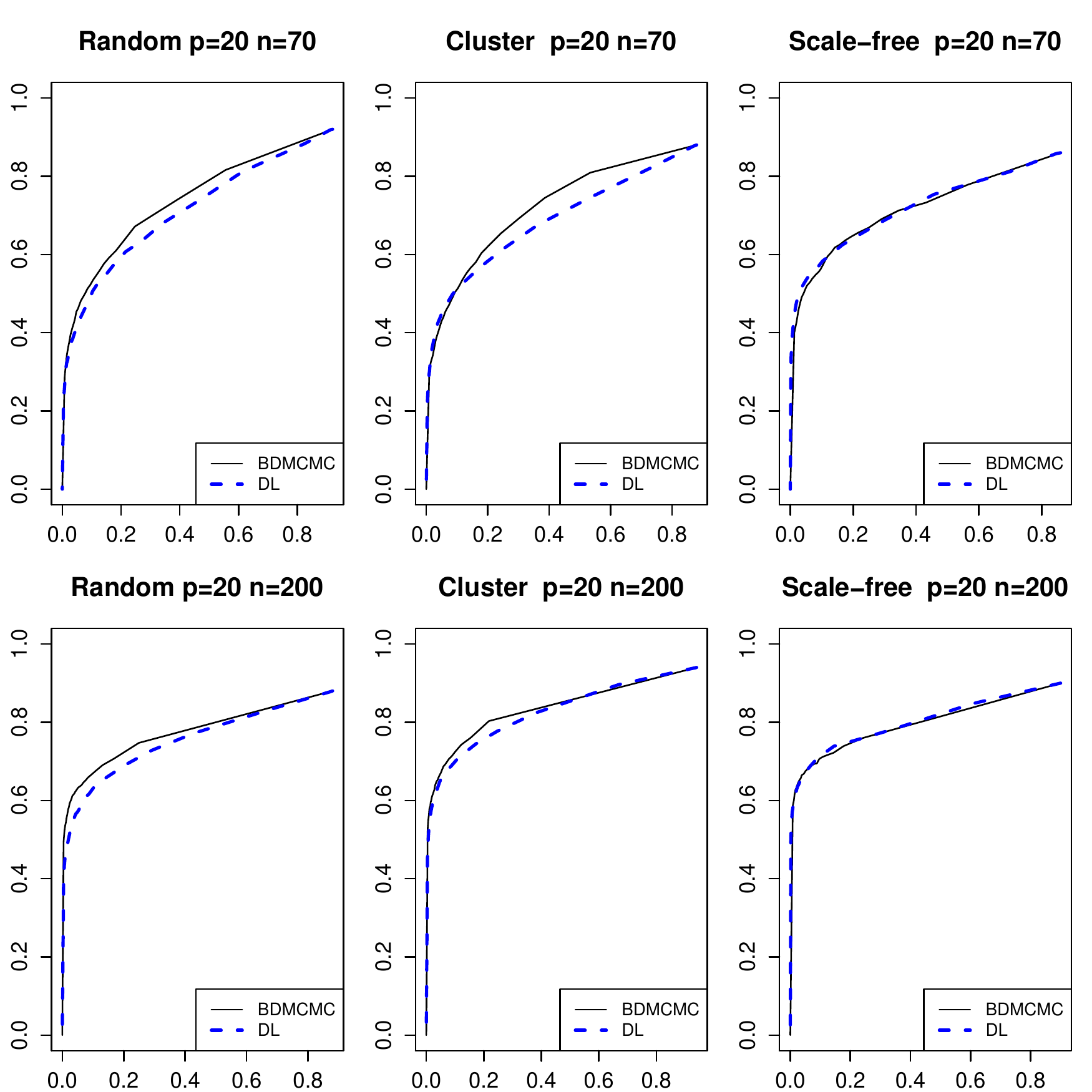}
\caption{ ROC curves depicting the performances of the simulated graphs for the case $p=20$ and $n=\{70,200\}$, based on the posterior edge inclusion probabilities. }
\label{fig:plotroc_p20}
\end{figure}

Another way to check the performance of EBDMCMC algorithm and compare it with DL is based on the ROC curves as shown in Figures \ref{fig:plotroc_p10} and \ref{fig:plotroc_p20}. %, \ref{fig:plotroc_p30}, and \ref{fig:plotroc_p40}.
The ROC curves depicting the performance of edge selection for each simulated graphs for the case $p=10$ and $n=10$, by varying thresholds on the posterior edge 
inclusion probabilities. As can be seen from the ROC curves the EBDMCMC performs better than the DL method.

%%%%%%%%%%%%%%%%%%%%%%%%%%%%%%%%%%%%%%%%%%%%%%%%%%%%%%%%%%%%%%%%%%%%%%%%%%%%%%%%%%%%%%%%%%%%%%%%%%%%%%%%%%%%%%%%%%%%%%%%%%%%%%%%%%%%%%%%%%%%%%%%%%%%%%%%%%%%%%%%%%%%%
\section{ Analysis of Dupuytren disease data }
\label{sec: dupuytren data}

\begin{figure}[ht]
\centering
\subfigure[Boxplot]{
   \centering
   \includegraphics[width=0.45\textwidth]{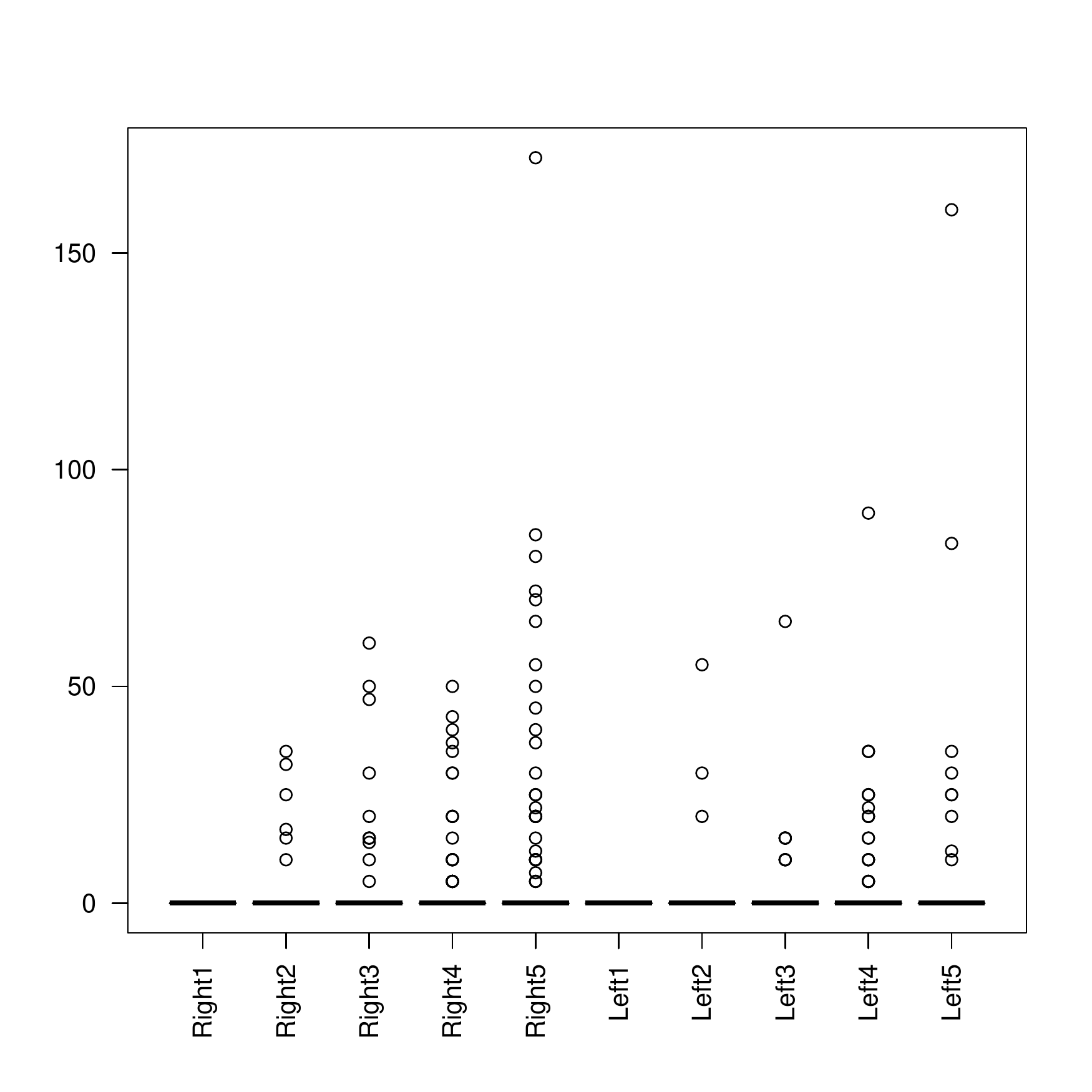}
   \label{fig:data boxplot}
}
\subfigure[Frequency histogram]{
   \centering
    \includegraphics[width=0.45\textwidth]{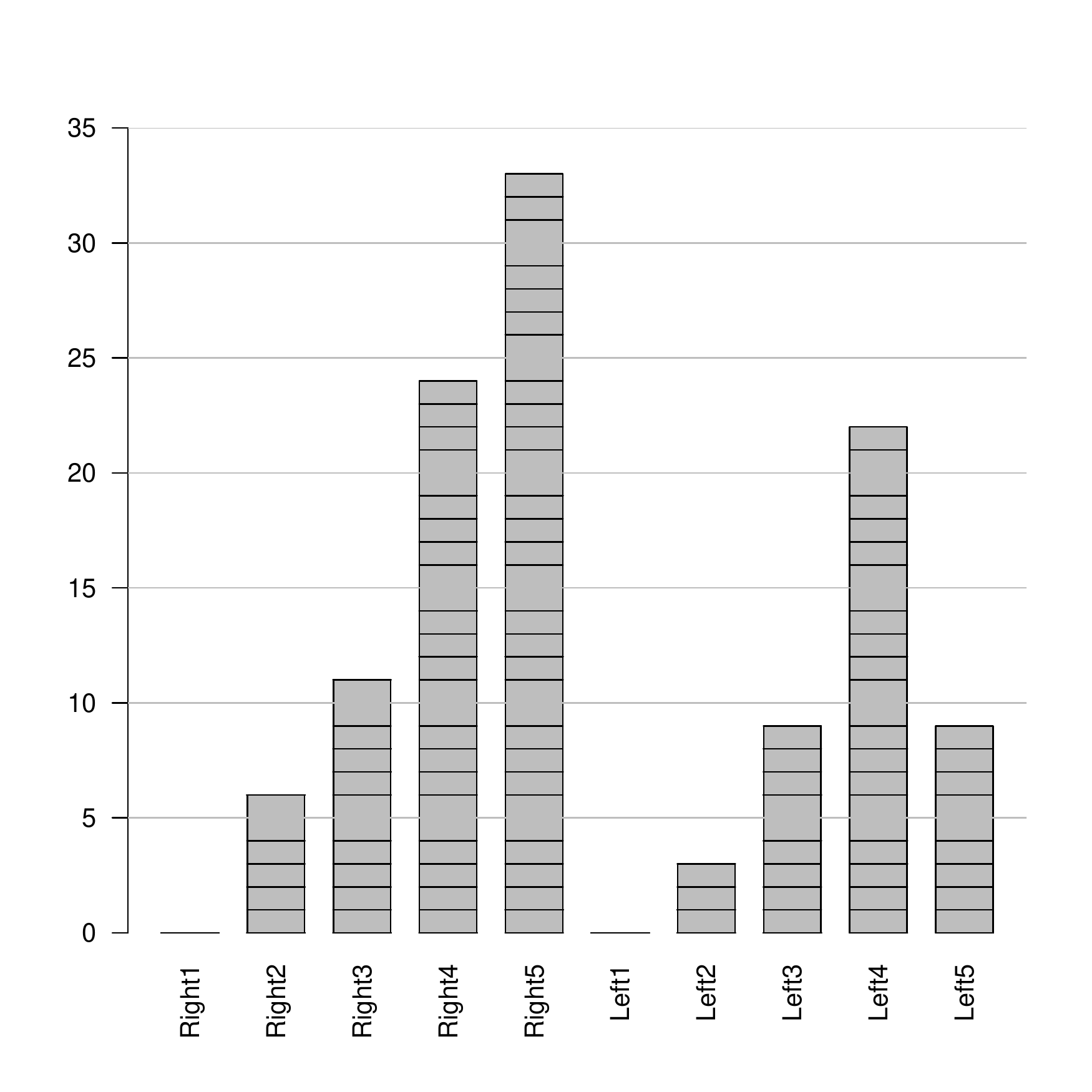}
    \label{fig:data hist}
}
\caption[]{ (a) Angles of the 10 fingers of all $279$ patients.
            (b) Frequency histogram of rays affected with Dupuytren disease for all 10 fingers. }
\label{fig:plot data}
\end{figure}

Here we analyze the data collected on patients who have Dupuytren disease in both hands from north of Netherlands 
by the Department of Plastic Surgery of the University Medical Center Groningen. 
The data are originally described by \cite{lanting2013prevalence} and \cite{lanting2014patterns}.
The data consist of $279$ patients who have Dupuytren disease ($n=279$); among those patients, $79$ of them 
have an irreversible flexion contracture in at least on one of their fingers. 
Therefore, the data consist of a lot of zeros as shown in Figure \ref{fig:plot data}. 
That is, though the hands, are affected by the disease, the fingers didn't show any sign of contraction and the total angle measure is taken as $0$.
% Besides, the data include $13$ potential risk factors.
% As a result we have mixed data that contains binary (disease factors), ordered discrete (alcohol and hand injury), and continuous variables (total angles for fingers).

The severity of the disease in all $10$ fingers of the patients is measured by the angles of each finger which is the sum of angles for metacarpophalangeal joints.
To study the potential phenotype risk factors of Dupuytren disease, we consider potential phenotype risk $13$ factors. 
These are smoking habits ({\bf Smoking}), alcohol consumption ({\bf Alcohol}), whether participants performed manual labour during a significant part of their life ({\bf Labour}), 
whether they had sustained hand injury in the past including surgery ({\bf HandInjury}), disease history information about the presence of {\bf Ledderhose}, {\bf Diabetes}, 
{\bf Epilepsy}, {\bf Peyronie}, {\bf Knuckle pad}, and liver disease ({\bf LiverDiseas}) and 
familial occurrence of Dupuytren disease which is defined as a first-degree relative with Dupuytren disease ({\bf Relative}).

For each finger we measure angles of metacarpophalangeal joints, two interphalangeal joints (for thumb fingers we only measure two interphalangeal joints); 
then we sum those angles for each fingers as a measure of the severity of Duptytren disease. 
The total angles ideally can vary from $0$ to $270$ degrees; however, in this dataset the minimum degree is $0$ and maximum $157$ degrees.
The age of participants (in years) ranges from $40$ to $89$ years, with an average age of $66$ years.
Smoking is binned into $3$ ordered categories (never, stopped, and smoking). 
Amount of alcohol consumption is binned into $8$ ordered categories (ranging from no alcohol to more than $20$ consumption per week). 
All other variables are binary.

% Phenotype risk factors previously described include alcohol consumption, smoking, manual labor, hand trauma, diabetes mellitus, and epilepsy 
% \citep{shih2010scientific, lanting2013prevalence}.
% \cite{lanting2014patterns} analyzes the Dupuytren disease with a multivariate ordinal logit model, taking into account age and sex, and tested hypotheses of the independence between groups of fingers.

In Section \ref{subsec: risk factors}, we infer the Dupuytren disease network with $13$ potential risk factors based on the EBDMCMC approach. 
In Section \ref{subsec: 10 fingers}, we consider only the severity measurements of the $10$ fingers to infer the interaction among the fingers.

%%%%%%%%%%%%%%%%%%%%%%%%%%%%%%%%%%%%%%%%%%%%%%%%%%%%%%%%%%%%%%%%%%%%%%%%%%%%%%%%%%%%%%%%%%%%%%%%%%%%%%%%%%%%%%%%%%%%%%%%%%%%%%%%%%%%%%%%%%%%%%%%%%%%%%%%%%%%%%%%%%%%%
\subsection{Inference for Dupuytren disease with risk factors}
\label{subsec: risk factors}

% We consider the severity of disease in all 10 hand fingers of the patients and $13$ potential phenotype risk factors of the Dupuytren disease, so we have a total of $p = 23$ variables.
% The potential risk factors are: age, sex, smoking, amount of alcohol (Alcohol), relative (Relative), number of hand injury of patients (HandInjury), Manual labour (Labour), 
% Ledderhose disease (Ledderhose), diabetes disease (Diabetes), epilepsy disease (Epilepsy), liver Disease (LiverDisease), peyronie disease (Peyronie), knuckle pad disease (Knucklepad). 
% For each finger we measure angles of metacarpophalangeal joints, two interphalangeal joints (for thumb fingers we only measure two interphalangeal joints); 
% then we sum those angles for each fingers as a measure of the severity of Duptytren disease. 
% The total angles ideally can vary from $0$ to $270$ degrees; however, in this dataset the minimum degree is $0$ and maximum $157$ degrees.
% The age of participants (in years) ranges from $40$ to $89$ years, with an average age of $66$ years.
% Smoking is binned into $3$ ordered categories (never, stopped, and smoking). 
% Amount of alcohol consumption is binned into $8$ ordered categories (ranging from no alcohol to more than $20$ consumption per week). 
% All other variables are binary.

\begin{figure} [!ht]
  \centering
      \includegraphics[width=0.85\textwidth]{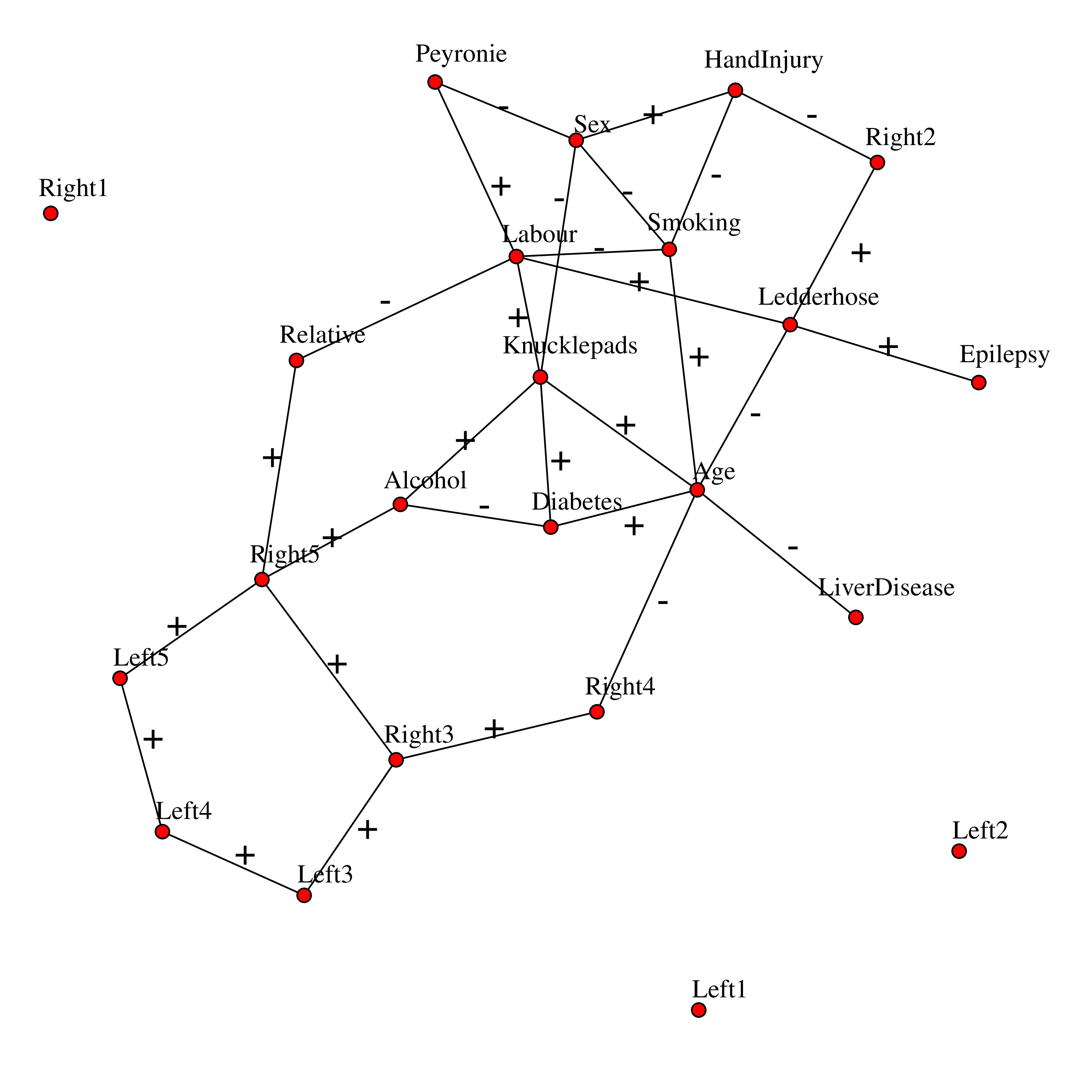}
  \caption{
    The inferred graph for the Dupuytren disease dataset based on $13$ risk factors and the total degrees of flexion in all 10 fingers. 
    It reports the selected graph with $26$ edges for which their posterior inclusion probabilities \eqref{posterior-edge} are more than $0.5$.
    Sign ``+'' shows a positive relationship between nodes and ``-'' shows a negative relationship.
    }
  \label{fig:graph-24-var-cut5}
\end{figure}

\begin{figure} [!ht]
  \centering
      \includegraphics[width=0.8\textwidth]{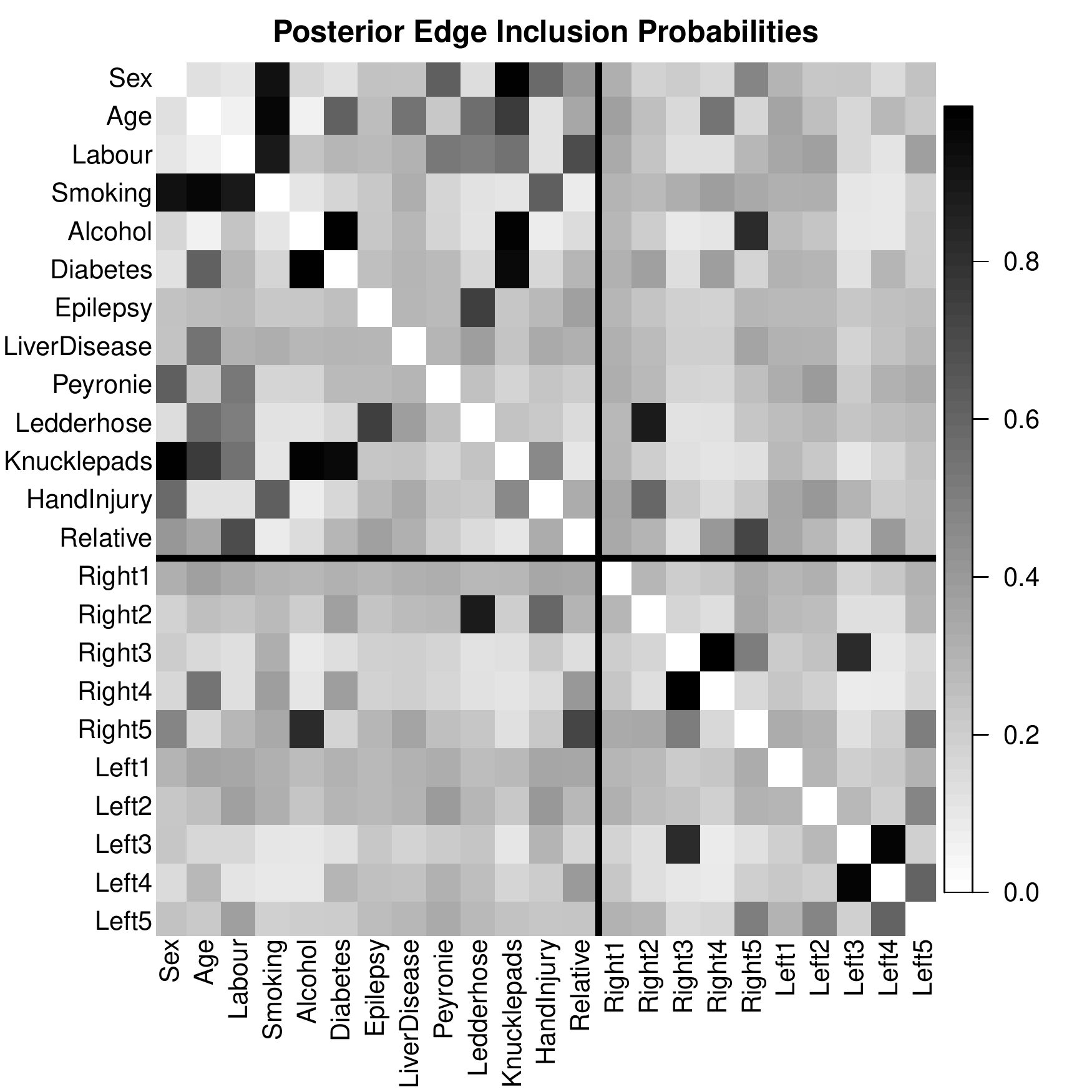}
  \caption{ Image visualization of the posterior edge inclusion probabilities of all possible edges in the graph, for 10 fingers with $13$ risk factors. }
  \label{fig:phat-image-24-variables}
\end{figure}

We apply our Bayesian framework to infer the conditional (in)dependence among the 23 variables and to identify the potential risk factors of Dupuytren disease and discover how they affect 
the disease. In implementing the proposed EBDMCMC approach to analyze this dataset, 
we place a uniform distribution as an uninformative prior on the graph and the G-Wishart prior $W_G(3,I_{23})$ on the precision matrix. 
We run the EBDMCMC algorithm for $2,000$K iterations with a $1,000$K sweeps burn-in. 
The results are displayed in Figures \ref{fig:graph-24-var-cut5} and \ref{fig:phat-image-24-variables}. 

The graph with the highest posterior probability is the graph with $42$ edges. 
% However, of those edges in the graph with highest posterior probabilities, basically, we 
Figure \ref{fig:graph-24-var-cut5} visualizes $26$ edges that have posterior probabilities large than $0.5$. 
Similarly, Figure \ref{fig:phat-image-24-variables} shows the image of all posterior inclusion probabilities where the degree of darkness increases with increasing posterior probabilities. %for visualization.

The edges in the graph show the interactions among the 10 severity measurements of Duputren disease and 13 risk factors. 
For example, the results (Figures \ref{fig:graph-24-var-cut5} and \ref{fig:phat-image-24-variables}) show that factors ``Age'', ``Alcohol'', ``Ledderhose disease'', 
``Hand Injury'' and ``Relative'',   among those $13$ risk factors,
have a significant association with the severity of Dupuytren disease. Graph \ref{fig:graph-24-var-cut5} also shows that factor ``Age'' 
is a hub in this graph and it plays a significant role as it affects the severity of the disease directly and
indirectly through the influence of other risk factors such as ``Ledderhose''.

\begin{figure} [!ht]
\centering
    \includegraphics[width=0.7\textwidth]{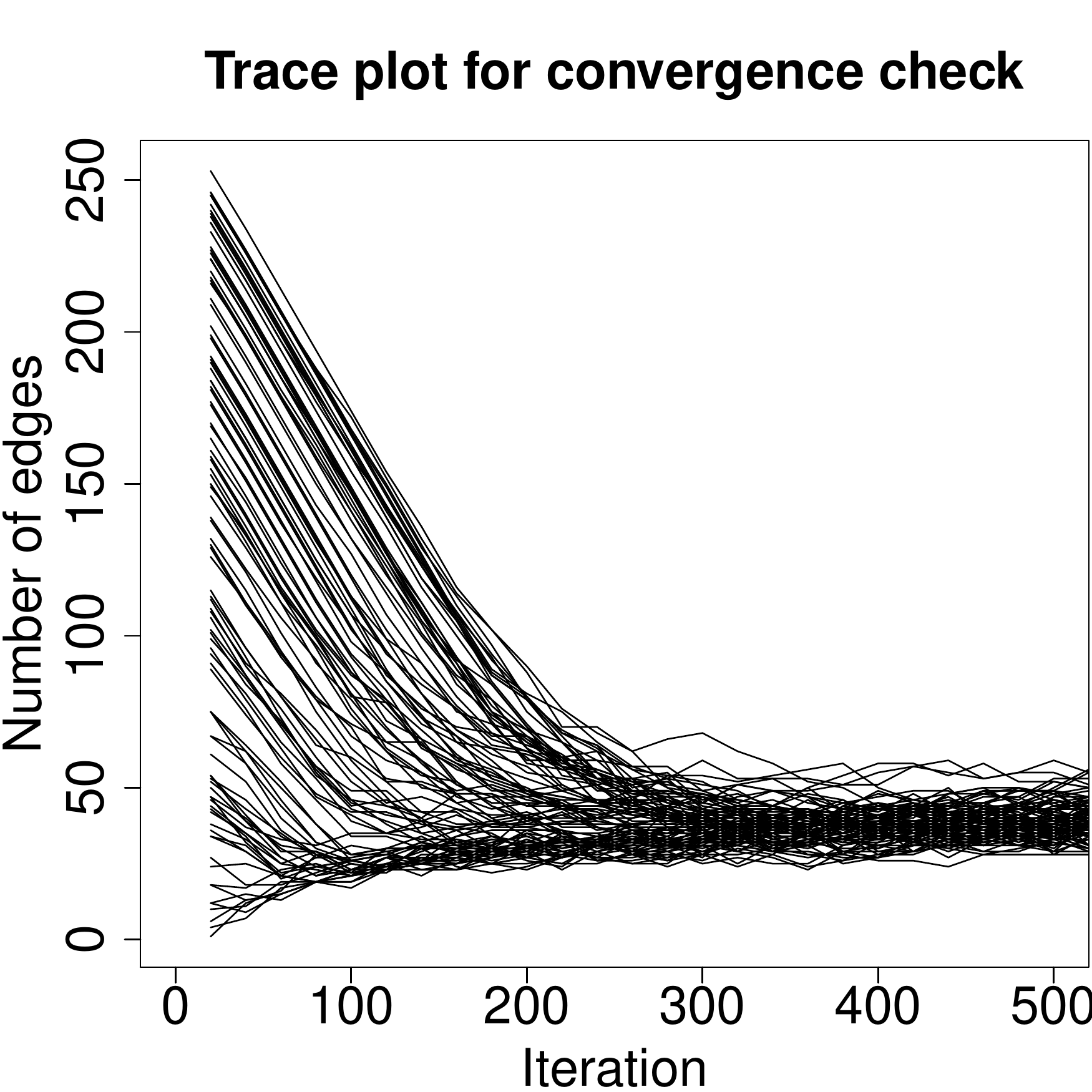}
\caption{ Case study of Section \ref{subsec: risk factors}. Trace plot of the number of edges included in the estimated graphs against iterations of the EBDMCMC algorithm with $100$ different starting points. } 
\label{fig:traceplat-24-var}
\end{figure}

Further we checked the stability of the selected graph with highest posterior probability at convergence of the algorithm with $100$ different starting points. 
The resulting Figure \ref{fig:traceplat-24-var} shows the traces of number of edges in the estimated graphs plotted against iterations of the EBDMCMC algorithm with the $100$ different starting points.
The plot shows good mixing around a stable graph model size which is $42$ and the algorithm converges after around $300$ iterations.

%%%%%%%%%%%%%%%%%%%%%%%%%%%%%%%%%%%%%%%%%%%%%%%%%%%%%%%%%%%%%%%%%%%%%%%%%%%%%%%%%%%%%%%%%%%%%%%%%%%%%%%%%%%%%%%%%%%%%%%%%%%%%%%%%%%%%%%%%%%%%%%%%%%%%%%%%%%%%%%%%%%%%
\subsection{Severity of Dupuytren disease between pairs of fingers}
\label{subsec: 10 fingers}

Here, we consider the relationship between incurrence of Dupuytren disease in pairs of fingers on both right and left hands. %of the10 hand fingers.  
Interaction between fingers is important because it help surgeons to decide whether they should operate one finger or multiple fingers simultaneously. 
The main idea is that if fingers are almost independent in terms of the severity of Dupuytren disease, there is no reason to operate the fingers simultaneously. 
On the other hand, if there is a strong relationship between fingers, then joint surgery may be recommended if one of the fingers is affected. 

\begin{figure} [!ht]
\centering
    \includegraphics[width=0.8\textwidth]{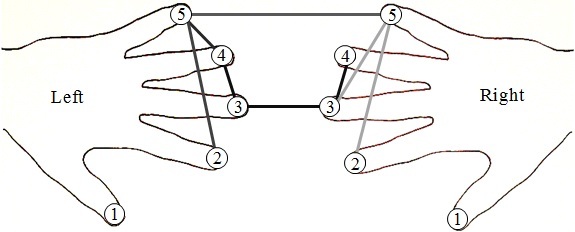}
\caption{
The inferred graph of Dupuytren disease dataset based on the total degrees of flexion in all 10 fingers. 
It reports the selected graph with $8$ edges for which their posterior inclusion probabilities \eqref{posterior-edge} are more than $0.5$.
}
\label{fig:grpah-fingers}
\end{figure}

\begin{figure} [!ht] 
\centering
    \includegraphics[width=0.7\textwidth]{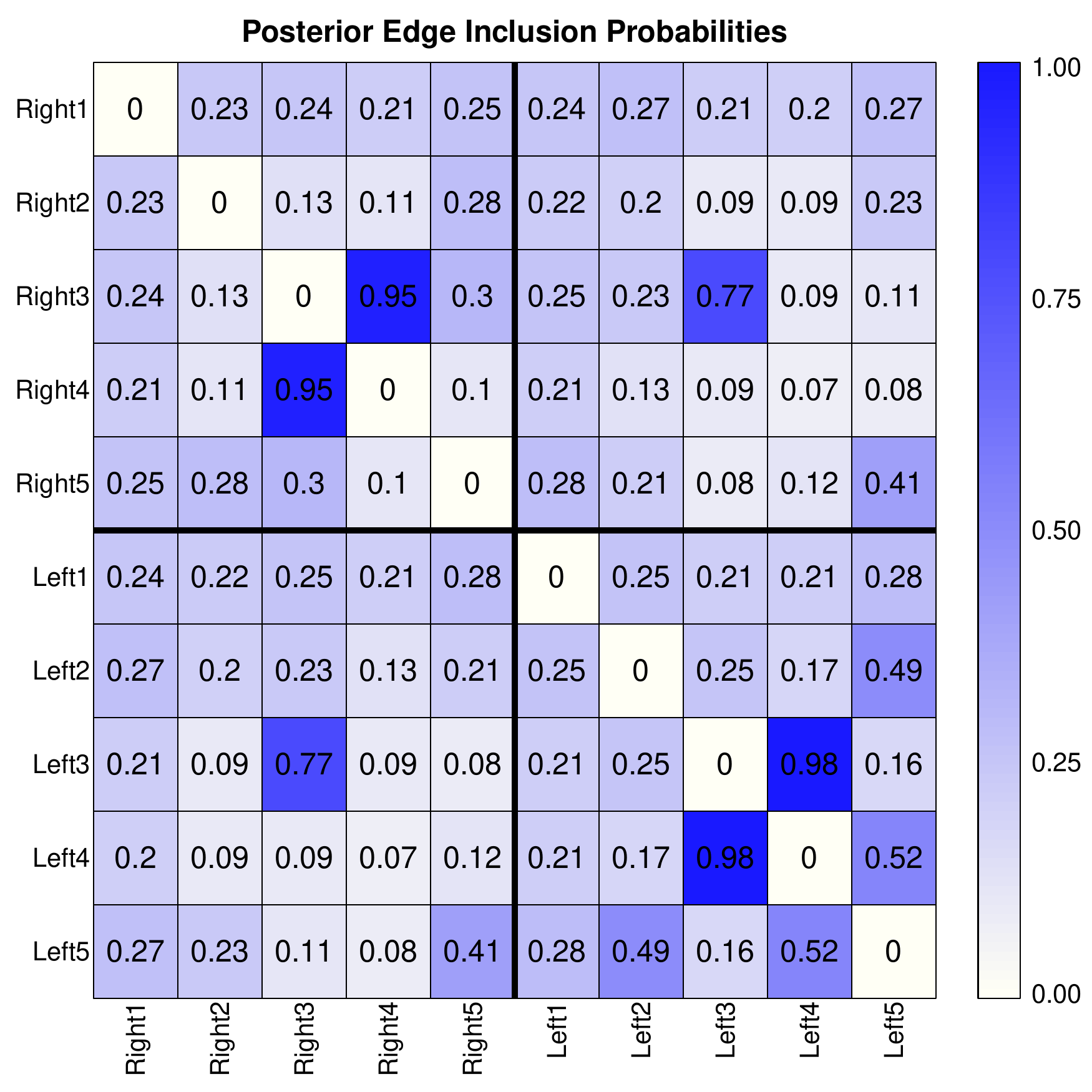}
\caption{ Image visualization of the posterior edge inclusion probabilities of all possible edges in the graph, for 10 fingers. }
\label{fig:phat-image-10-fingeres}
\end{figure}

We apply the proposed EBDMCMC approach for the $10$ variables of Dupuytren disease severity measures using a uniform distribution as an uninformative prior on the graph and the G-Wishart $W_G(3,I_{10})$ prior on the precision matrix. 
We run the EBDMCMC algorithm for $2,000$K iterations with a $1,000$K as burn-in. The results are displayed in Figure \ref{fig:grpah-fingers} and Figure \ref{fig:phat-image-10-fingeres}.

The graph with the highest posterior probability is the graph with $12$ edges. 
Figure \ref{fig:grpah-fingers} visualizes the selected graph with $8$ edges, for which the posterior inclusion probabilities in \eqref{posterior-edge} are greater than $0.5$.
The edges in the graph show the interactions among the fingers with regard to the severity of Dupuytren disease.

The results (Figures \ref{fig:grpah-fingers} and \ref{fig:phat-image-10-fingeres}) show significant co-occurrences of Dupuytren disease in the ring fingers and middle fingers in both hands. 
This suggests that presence of disease in the middle finger is strongly associated to the ulnar side of the hand. 
Surprisingly, our results also show a strong relationship between middle fingers in both hands.
% There is also a strong association between the middle finger and ring finger on both right and left hands. 
Moreover, the results show that the joint interactions between fingers in both hands are almost symmetric.
These results are in support of the hypotheses that the disease has genetic factors or other biological factors that affect similar fingers in both hands %fingers simultaneously.

%%%%%%%%%%%%%%%%%%%%%%%%%%%%%%%%%%%%%%%%%%%%%%%%%%%%%%%%%%%%%%%%%%%%%%%%%%%%%%%%%%%%%%%%%%%%%%%%%%%%%%%%%%%%%%%%%%%%%%%%%%%%%%%%%%%%%%%%%%%%%%%%%%%%%%%%%%%%%%%%%%%%%
\subsection{Fit of model to Dupuytren data}
\label{subsec: Model checking}

Posterior predictive checks can be used for checking whether the proposed Bayesian approach fits the Dupuytren data set well or not. 
If the model fits the Dupuytren data, then simulated data generated under the model should look like to the observed data.
In this regard, first, based on our estimated graph from the EBDMCMC algorithm in section \ref{subsec: risk factors}, 
we draw simulated data from the posterior predictive distribution. % of replicated data.
Then, we compare the samples to our observed data.
Any systematic differences between the simulations and the data determine potential failings of the model.

\begin{figure} [!ht]
\centering
    \includegraphics[width=0.8\textwidth]{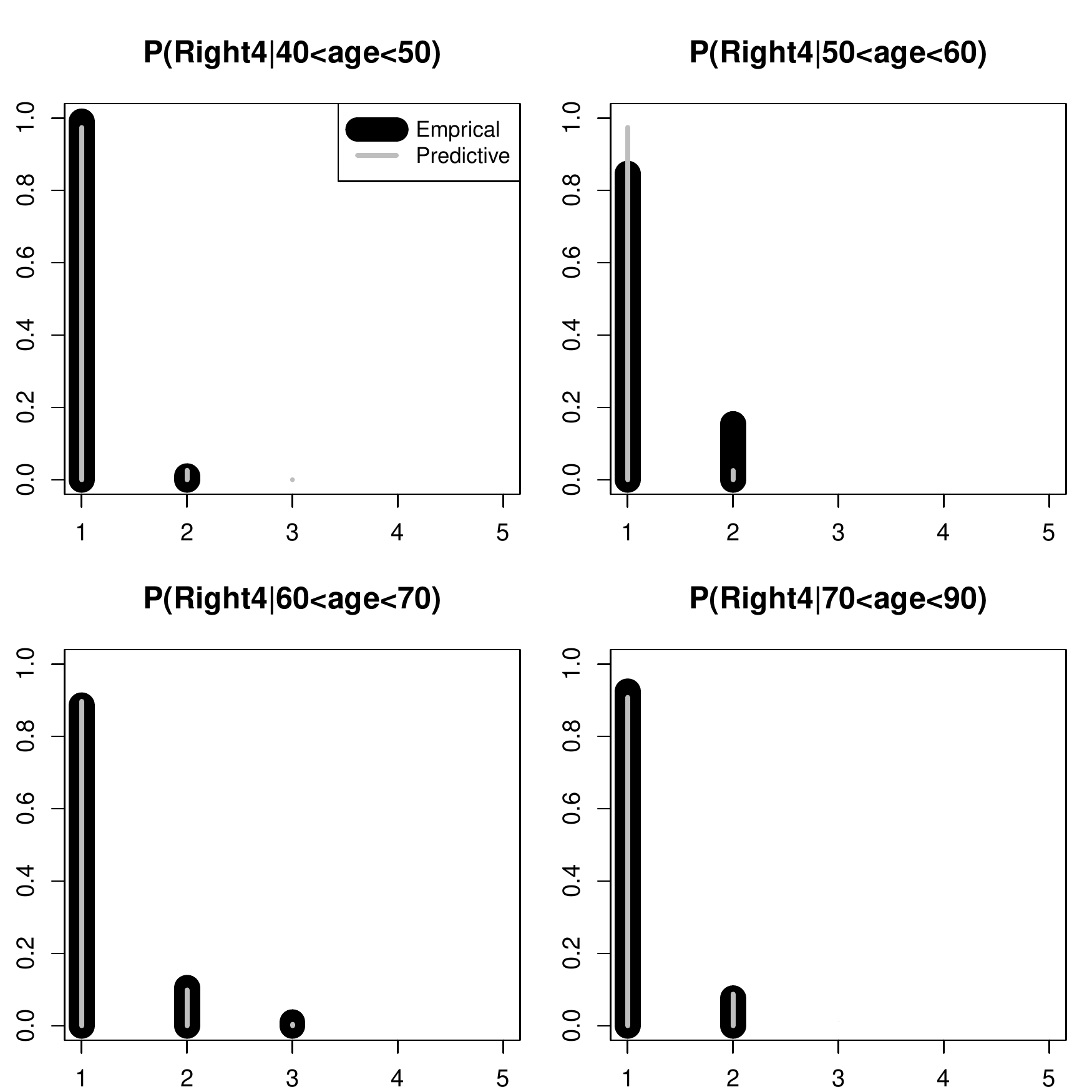}
\caption{ Empirical and predictive conditional distributions for total angle of finger 4 in right hand condition on four different categories of variable \textquotedblleft age\textquotedblright. }
\label{fig:age-right4}
\end{figure}

\begin{figure} [!ht]
\centering
\begin{minipage}[b]{0.45\linewidth}
 \centering
 \includegraphics[width=0.8\textwidth]{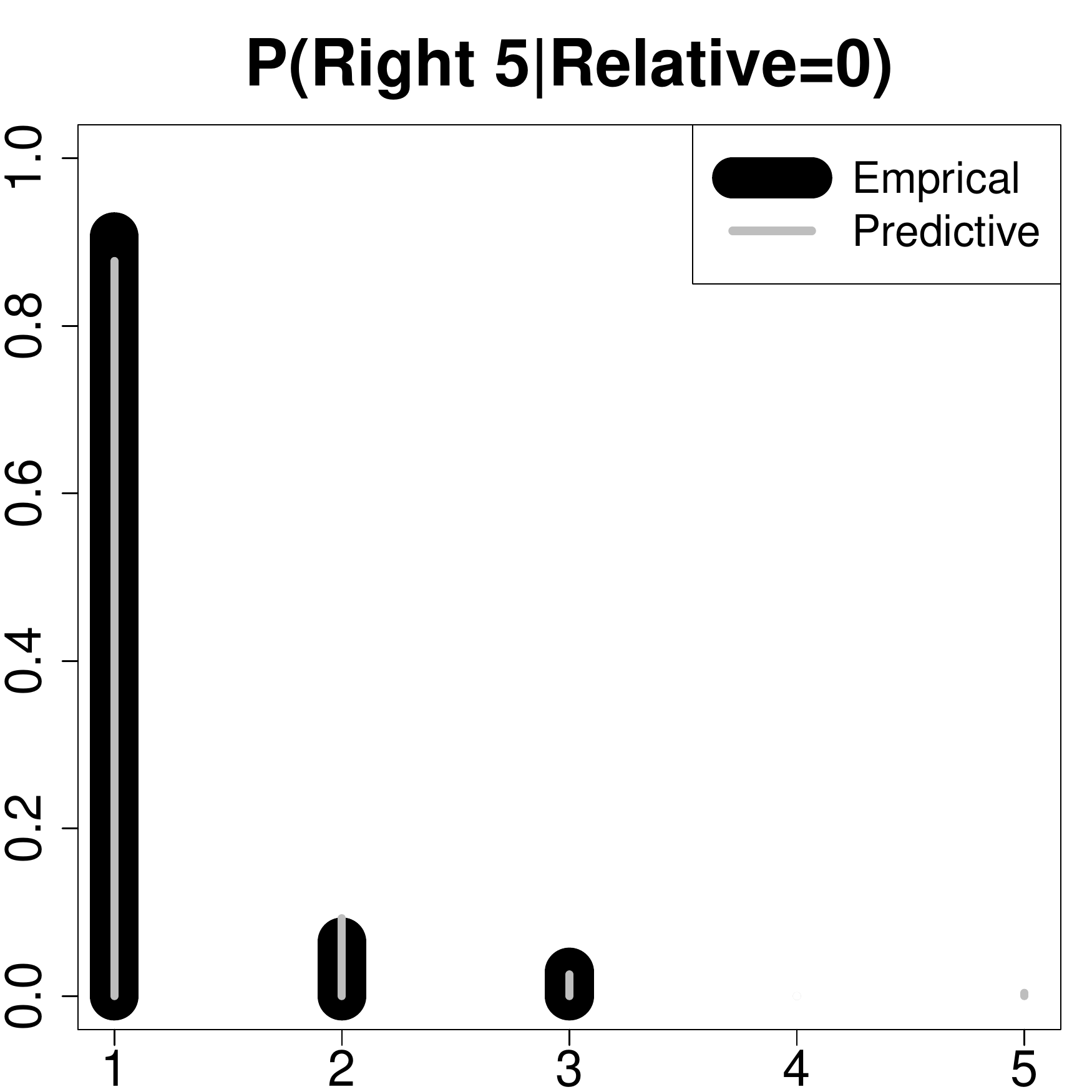}
\end{minipage}
\hspace{0cm}
\begin{minipage}[b]{0.45\linewidth}
 \centering
 \includegraphics[width=0.8\textwidth]{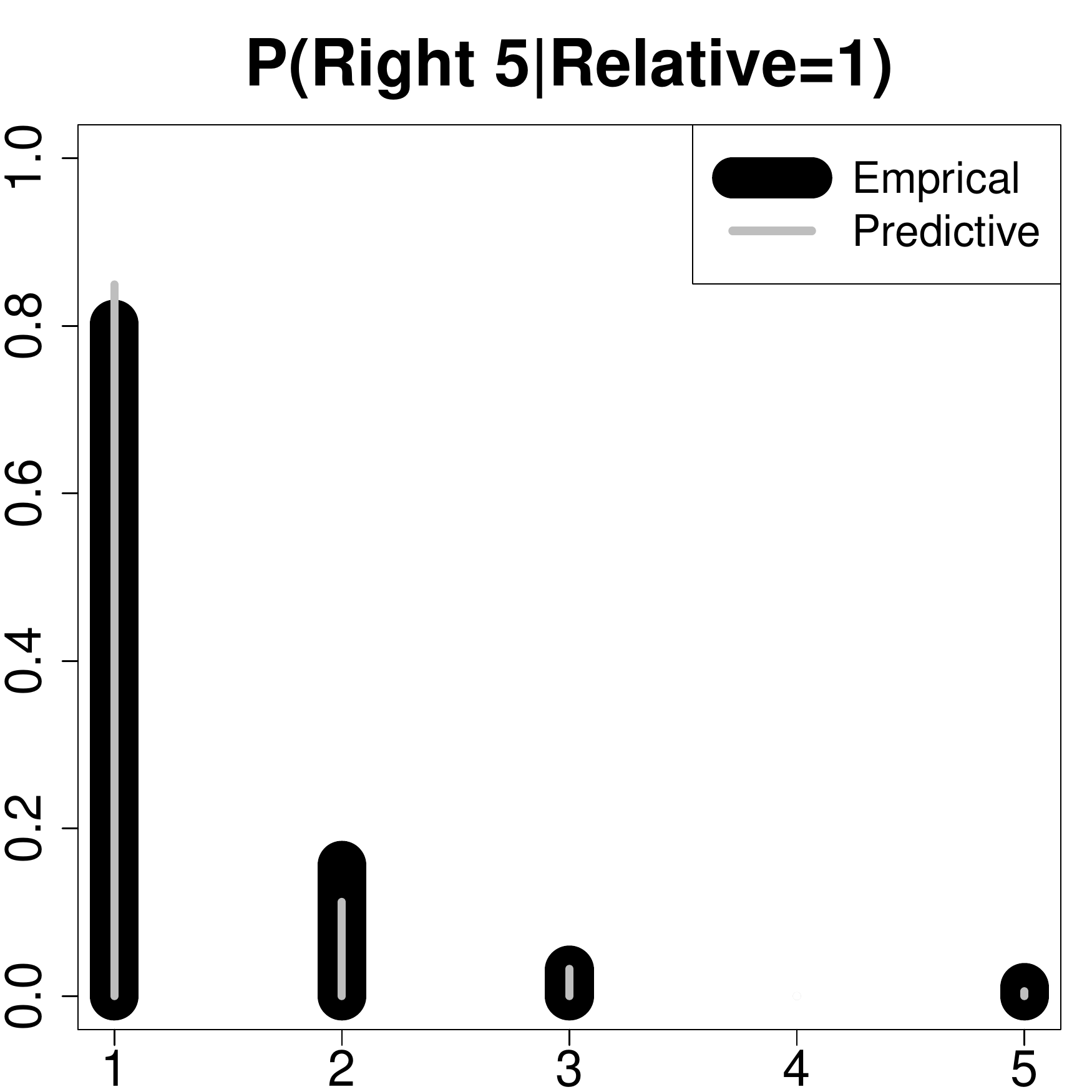}
\end{minipage}
\caption{ Empirical and predictive conditional distributions for total angles of finger 5 in right hand condition on relative variable. }
 \label{fig:right5-relative}
\end{figure}

\begin{figure}[!ht]
\centering
\begin{minipage}[b]{0.45\linewidth}
 \centering
 \includegraphics[width=0.8\textwidth]{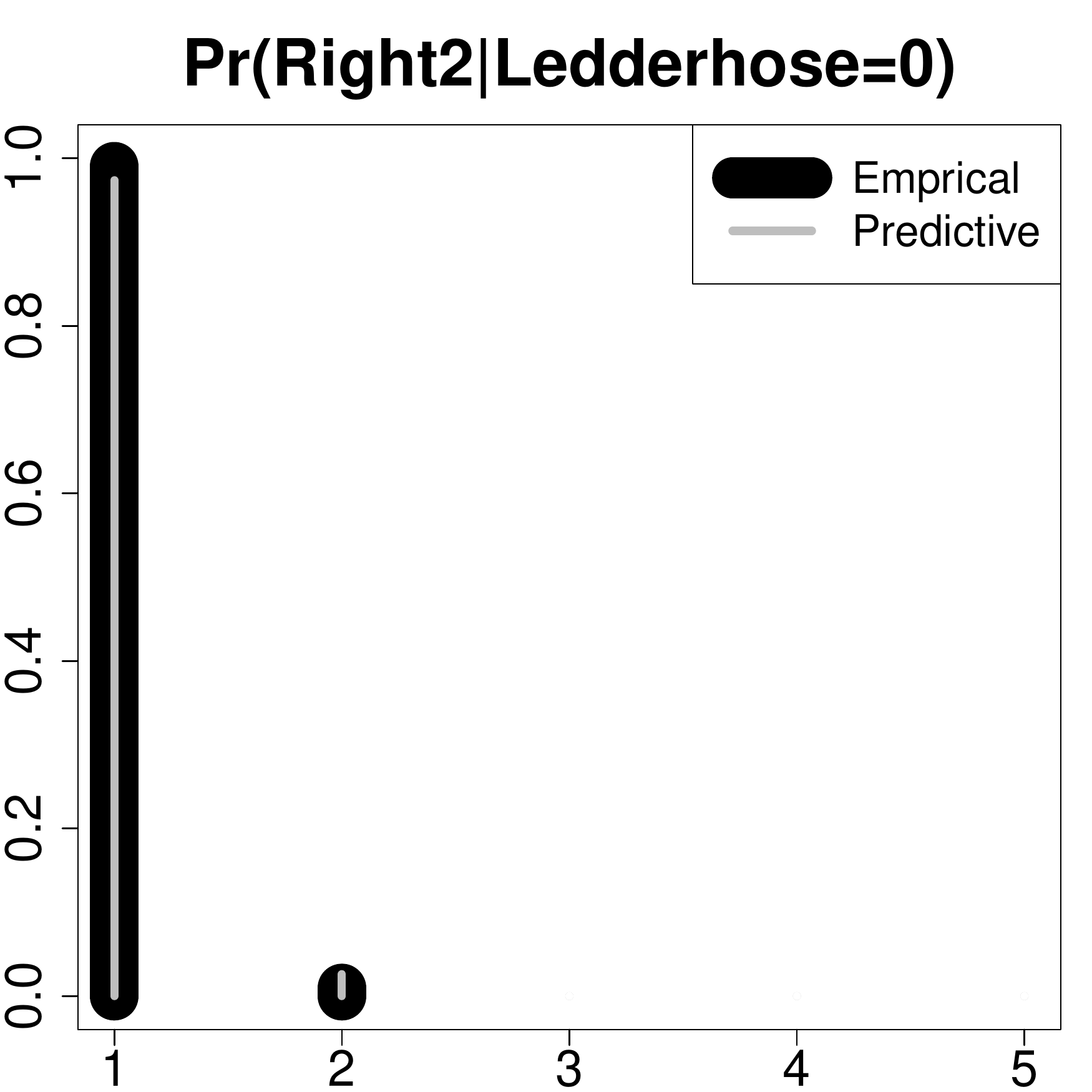}
\end{minipage}
 \hspace{0cm}
\begin{minipage}[b]{0.45\linewidth}
 \centering
 \includegraphics[width=0.8\textwidth]{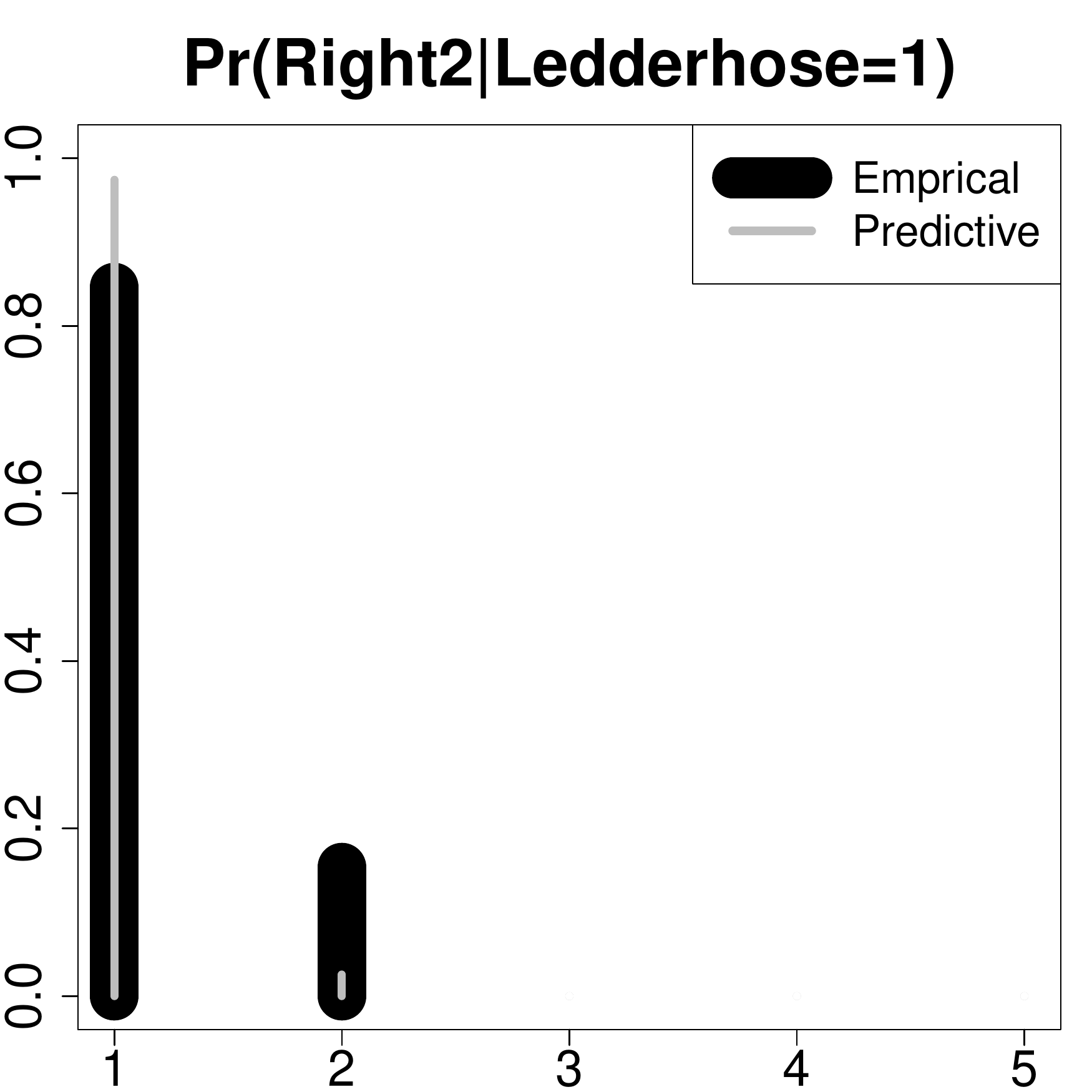}
\end{minipage}
\caption{Empirical and predictive conditional distributions for total angles of finger 2 in right hand condition on Ledderhose disease variable.}
 \label{fig:right2-Ledderhose}
\end{figure}

We obtain the conditional distributions of the potential risk factors and disease severity measures on the fingers for both simulated and observed data. 
The empirical and predictive conditional distributions of some selected variables are presented in Figures \ref{fig:age-right4}, \ref{fig:right5-relative} and \ref{fig:right2-Ledderhose}.

Figure \ref{fig:age-right4} displays the empirical and predictive distributions of disease severity measure on finger 4 in right hand (right4) conditional on variable 
\textquotedblleft age\textquotedblright in four categories $\{ (40, 50)$, $(50, 60)$, $(60, 70)$, $(70, 90) \}$. 
The variable right4, based on Tubiana Classification, grouped into $5$ categories, 
category $1$: $0$ degree for total angle; $2$: degree between $(1, 45)$; $3$: degree between $(46, 90)$; $4$: degree between $(90, 135)$; $5$: degree more than $135$.
 The results in Figure \ref{fig:age-right4} show that the fit is good, since the predicted conditional distributions, in general, are the same with the empirical distributions. 

Similarly, Figure \ref{fig:right5-relative} plots the empirical and predictive distribution of disease severity measure on finger 5 in right hand (right5) 
conditional on variable \textquotedblleft Relative\textquotedblright and Figure \ref{fig:right2-Ledderhose} plots the empirical and predictive distribution of 
disease severity measure on finger 2 in right hand (right2) conditional on variable \textquotedblleft Ledderhose\textquotedblright. 
These results also suggest that the EBDMCMC algorithm fits the Dupuytren data well as the predicted conditional distributions are in agreement with the empirical distributions.

%%%%%%%%%%%%%%%%%%%%%%%%%%%%%%%%%%%%%%%%%%%%%%%%%%%%%%%%%%%%%%%%%%%%%%%%%%%%%%%%%%%%%%%%%%%%%%%%%%%%%%%%%%%%%%%%%%%%%%%%%%%%%%%%%%%%%%%%%%%%%%%%%%%%%%%%%%%%%%%%%%%%%
\section{Conclusion}
\label{sec: conclusion}

In this paper we have implemented a Bayesian method for discovering the effect of mixed potential risk factors of Dupuytren disease  %the effect of potential risk factors of Dupuytren disease 
and the
underling relationships between fingers on both right and left hands with regard to severity of the Dupuytren disease.

The results of the case study clearly demonstrate that age, alcohol, relative, and ledderhose diseases all affect Dupuytren disease directly.
Other risk factors only affect Dupuytren disease indirectly.
Another important result is that severity of Dupuytren disease in fingers are correlated.
In particular, the middle finger with the ring finger.
This implies that a surgical intervention on either the ring or middle finger should preferably be executed simultaneously.

In Section \ref{sec: dupuytren data}, based on our Bayesian framework, we model the Dupuytren disease by consider the $13$ potential risk factors. 
Our result in this section support the hypotheses the disease has genetic factors or other biological factors that affect the severity of the disease in fingers simultaneously.
Indeed, in genome-wide association study, nine genes were identified to be associated with Dupuytren disease \citep{dolmans2011wnt}. 
It should be interesting to consider the potential environmental risk factors jointly with those biological factors.
Bayesian inference for all these factors requires a computationally efficient search algorithm that can potentially explore the underlying graph structure to uncover 
complicated patterns among these variables. 
% In this regard, the proposed EBDMCMC approach can deal with such a high-dimensional problems; we show it in our simulation example.
% Considering all those phenotype and genotype based on our Bayesian framework will be studied in future work.

We compare our EBDMCMC Bayesian approach with an alternative Bayesian approach \citep{dobra2011copula} using a simulation study on various types of graph structures. 
Although, both approaches converge to the same posterior distribution our approach has some clear advantages on finite MCMC runs. 
This difference is mainly due to our implementation of a computationally efficient algorithm. Our method is computationally more efficient because of two reasons.
Firstly, our sampling algorithm is based on birth-death process.
% which compare to the RJMCMC implemented in \cite{dobra2011copula} is much more efficient. 
Secondly, we implemented an exact way of computing for the ratio of normalizing constants based on the result in \cite{uhler2014exact}, which has been computationally a bottle-neck in the Bayesian approach.

Of course, our extended Bayesian method is not limited only to this type of data.
It can potentially be applied to any kind of mixed data where the observed variables are binary, ordinal or continuous.

%%%%%%%%%%%%%%%%%%%%%%%%%%%%%%%%%%%%%%%%%%%%%%%%%%%%%%%%%%%%%%%%%%%%%%%%%%%%%%%%%%%%%%%%%%%%%%%%%%%%%%%%%%%%%%%%%%%%%%%%%%%%%%%%%%%%%%%%%%%%%%%%%%%%%%%%%%%%%%%%%%%%%%%%%%%%%%%%%%
% references
\bibliographystyle{Chicago}
\bibliography{ref}

%%%%%%%%%%%%%%%%%%%%%%%%%%%%%%%%%%%%%%%%%%%%%%%%%%%%%%%%%%%%%%%%%%%%%%%%%%%%%%%%%%%%%%%%%%%%%%%%%%%%%%%%%%%%%%%%%%%%%%%%%%%%%%%%%%%%%%%%%%%%%%%%%%%%%%%%%%%%%%%%%%%%%%%%%%%%%%%%%%
% end for document
\end{document}